\documentclass[lettersize,journal]{IEEEtran}
\usepackage{amsmath,amsfonts}
\usepackage{algorithm}

\usepackage{algpseudocode}
\usepackage{booktabs}
\usepackage{caption}
\usepackage{lineno}
\usepackage{multirow}
\usepackage{makecell}

\usepackage[table]{xcolor}
\usepackage[margin=1in]{geometry}
\usepackage{tabularx}
\usepackage{bm}
\usepackage[unicode=true]{hyperref} 
\usepackage{cleveref}
\usepackage{subcaption} 
\definecolor{customorange}{RGB}{218,142,0}
\definecolor{customred}{RGB}{219,83,75}

\usepackage{array}

\usepackage{textcomp}
\usepackage{stfloats}
\usepackage{url}
\usepackage{verbatim}
\usepackage{graphicx}
\usepackage{cite}
\newcommand{\REQUIRE}{\textbf{Require: }}
\begin{document}

\title{T3MAL: Test-Time Fast Adaptation for Robust Multi-Scale Information Diffusion Prediction}

%

\author{Wenting Zhu, Chaozhuo Li, Qingpo Yang, Xi Zhang and Philip S. Yu~\IEEEmembership{Fellow,~IEEE}
        
\thanks{Manuscript received 16 July, 2025; revised 16 July, 2025.
\textit{(Corresponding author: Xi Zhang)}}

\thanks{Wenting Zhu, Chaozhuo Li, Qingpo Yang and Xi Zhang are with Beijing University of Posts and Telecommunications, Beijing 100876, China. Email:\{zwt, lichaozhuo, 2023111098, zhangx\}@bupt.edu.cn.}
\thanks{Philip s. Yu is with the University of Illinois at Chicago, Chicago, IL 60607, USA. E-mail: psyu@uic.edu.}
}
\markboth{Journal of \LaTeX\ Class Files,~Vol.~14, No.~8, August~2021}%
{Shell \MakeLowercase{\textit{et al.}}: A Sample Article Using IEEEtran.cls for IEEE Journals}


\maketitle

\begin{abstract}
Information diffusion prediction (IDP) is a pivotal task for understanding how information propagates among users. Most existing methods commonly adhere to a conventional “training-test” paradigm, where models are pretrained on training data and then directly applied to test samples. However, the success of this paradigm hinges on the assumption that the data are independently and identically distributed, which often fails in practical social networks due to the inherent uncertainty and variability of user behavior. In the paper, we address the novel challenge of distribution shifts within IDP tasks and propose a robust test-time training (TTT)-based framework for multi-scale diffusion prediction, named \textit{T3MAL}. The core idea is to flexibly adapt a trained model to accommodate the distribution of each test instance before making predictions via a self-supervised auxiliary task. Specifically, T3MAL introduces a BYOL-inspired self-supervised auxiliary network that shares a common feature extraction backbone with the primary diffusion prediction network to guide instance-specific adaptation during testing. Furthermore, T3MAL enables fast and accurate test-time adaptation by incorporating a novel meta-auxiliary learning scheme and a lightweight adaptor, which together provide better weight initialization for TTT and mitigate catastrophic forgetting. Extensive experiments on three public datasets demonstrate that T3MAL outperforms various state-of-the-art methods. 



\end{abstract}

%







\section{Introduction}
\IEEEPARstart{A}{s} large-scale information cascades are recorded on social platforms, researchers have been motivated to explore the underlying patterns of information diffusion. Current research in information diffusion prediction (IDP) primarily focuses on two key tasks: macroscopic prediction, which estimates the future popularity of a cascade, and microscopic prediction, which identifies the next potential retweeter. 

\begin{figure}[t]
\centering
\includegraphics[trim=1.5 0 3.5 0, clip, width=\columnwidth]{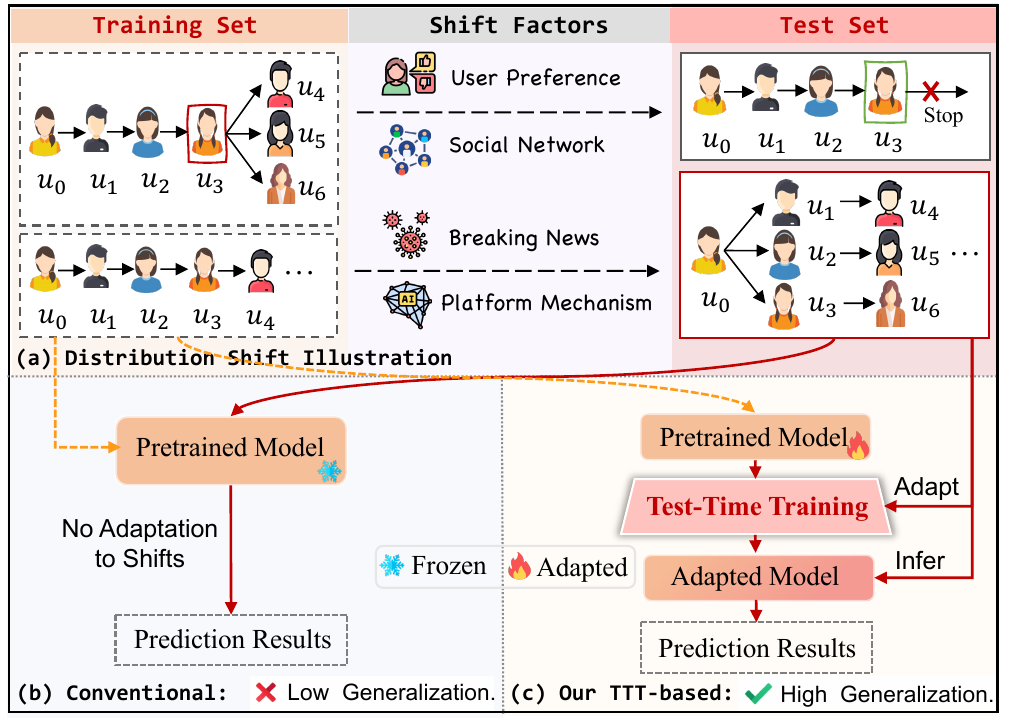}
\caption{(a) Illustration of distribution shifts in IDP tasks, highlighting key factors driving these shifts. (b) and (c) Comparison of conventional training methods and the proposed TTT-based approach. Orange dashed arrows indicate training, while red solid arrows indicate testing.}

\label{Intro fig}
\vspace{-2mm}  
\end{figure}

Previous studies primarily focus on either microscopic or macroscopic prediction tasks in isolation~\cite{xu2021casflow, sun2022ms, feng2022h, bao2024popularity}, whereas recent advances propose to integrate both tasks into a unified framework for multi-scale prediction~\cite{yang2021full, jiao2024enhancing}. Both existing single-scale and multi-scale prediction methods adhere to the conventional “training-test” paradigm, in which models are trained on the training set and then directly deployed to the test data, as illustrated in Fig.~\ref{Intro fig}(b). The success of such paradigm relies heavily on the independent and identically distributed (i.i.d.) assumption, under which training and testing cascades are sampled from the same distribution \cite{quinonero2022dataset}. 


Nevertheless, this assumption often fails to hold in real-world social platforms \cite{jia2022hedan}. As shown in Fig.~\ref{Intro fig}(a), diffusion patterns at test time may diverge significantly from those observed during training due to various factors, such as user preferences, social networks, and platform mechanisms \cite{yuan2021dyhgcn, wang2023exploring}. These factors evolve and interact in unpredictable ways, inevitably introducing uncertainty into user behavior and cascade trajectories. For example, the disengagement of an influential user $u_3$ from a community may interrupt downstream diffusion, leading to lower popularity \cite{ji2023community}, while a post by an epidemiologist $u_0$ may surge in popularity during the COVID-19 pandemic due to heightened public attention \cite{ziakas2024public,bao2024popularity}.
Such distribution gap between training and test cascades hinders model generalization and results in unsatisfactory prediction performance on test cascades.




In this paper, we investigate the critical yet underexplored problem of \textit{\textbf{distribution
shifts}} in IDP tasks. A common line of research addresses this issue by leveraging existing techniques from unsupervised domain adaptation (UDA) \cite{fernando2013unsupervised, long2018conditional, ganin2016domain} and domain generalization (DG) \cite{balaji2018metareg, carlucci2019domain, wang2022generalizing}. Both UDA and DG methods attempt to bridge the distribution gap between source and target domains by learning domain-invariant features during \textit{training}, with the expectation that the trained model will generalize well across arbitrary test cascades. However, test cascades often exhibit distinct properties, and the learned domain-invariant features may be insufficient to represent their unique patterns, limiting the model’s performance on specific instances. Moreover, these approaches often overlook the potential utility of unlabeled test data beyond evaluation, despite its implicit clues about the test distribution.

Inspired by recent advancements in Test-Time Training (TTT) in computer vision \cite{sun2020test, chen2022ost, Sain_2022_CVPR}, a more promising solution lies in dynamically adapting the trained model to the distribution of each newly encountered test cascade, rather than anticipating all possible distribution shifts during training. Following this insight, we aim to develop a robust TTT-based approach for multi-scale information diffusion prediction, without requiring any manual labels. As shown in Fig.~\ref{Intro fig}(c), our method is capable of performing instance-specific adaptation before making predictions at \textit{test time}, allowing the model to better generalize to unseen test cascades.

However, designing an effective TTT framework for IDP tasks is non-trivial due to several challenges. First, given the supervision-free nature of TTT, selecting an appropriate self-supervised auxiliary task is crucial for guiding the adaptation process at test time \cite{gandelsman2022test}. This task must be sufficiently general to extract useful features for the primary tasks across various distributions. Second, naively using the jointly trained model as the initialization for TTT, as done in vanilla TTT methods\cite{zhu2025ghidorah}, may be suboptimal, as it lacks foresight into future learning objectives—namely, instance-specific adaptation. Third, since the model adaptation at test time relies entirely on the auxiliary task, directly updating the trained model may distort the learned representation space and lead to \textit{catastrophic forgetting} of diffusion prediction knowledge \cite{mccloskey1989catastrophic}, resulting in degraded performance on the primary tasks (see detailed analysis in Section~\ref{subsec:analysis of catastrophic forgetting}). 





To tackle the aforementioned challenges, we propose \textbf{T3MAL}, a robust multi-scale diffusion prediction framework that integrates \textbf{\underline{TTT}} with a novel \textbf{\underline{M}}eta-\textbf{\underline{A}}uxiliary \textbf{\underline{L}}earning scheme to enable instance-specific adaptation under distribution shifts. Following the TTT paradigm, T3MAL consists of two key networks: a primary network for diffusion prediction and an auxiliary network for self-supervised learning, both sharing a common feature extraction backbone but with distinct prediction heads. For the auxiliary network, we empirically adopt BYOL \cite{grill2020bootstrap}, which has proven effective for learning informative feature representations for specific instances within our framework. To overcome the limitations of vanilla TTTs (i.e. catastrophic forgetting and suboptimal initialization), we introduce an additional meta-auxiliary learning phase after joint training. This phase simulates the TTT process via nested optimization while ensuring that the adaptation guided by the auxiliary task consistently benefits the primary tasks. Compared to the jointly trained model, the meta-trained model provides a better initialization for TTT, enabling fast and accurate test-time adaptation with only a few gradient steps.  
Moreover, to mitigate catastrophic forgetting, we freeze the feature encoder after joint training and design a lightweight adaptor that customizes the encoder for each cascade to better capture its unique characteristics. Evaluations on three datasets demonstrate that T3MAL significantly outperforms state-of-the-art models.







Our main contributions can be summarized as follows:

\begin{itemize}
\item We propose a novel TTT-based framework for robust multi-scale diffusion prediction under distribution shifts, which can flexibly adapt a trained model to each test cascade without requiring additional labels during inference. To the best of our knowledge, this is the first work to introduce the TTT paradigm to address the critical yet overlooked problem of distribution shifts in IDP tasks.

\item To overcome the limitations of vanilla TTT methods, we propose several innovative mechanisms, including a novel meta-auxiliary learning scheme and a lightweight adaptor. By learning a better weight initialization for TTT and mitigating catastrophic forgetting, our method enables fast and accurate test-time adaptation.
%
\item Extensive experiments on three public real-world datasets demonstrate that our framework consistently outperforms state-of-the-art methods.
\end{itemize} 

\section{PRELIMINARIES}
This section first formally defines the data structure we use and the IDP problems we are addressing. Then, we introduce the vanilla test-time training paradigm adopted in this work. Notations used throughout the paper are summarized in \textit{\textbf{Appendix A}} for reference.
\vspace{-0.8em}

\subsection{Problem Statement}
Since information spreads among social users in a cascading manner, we define the input as a set of cascades $\mathcal{C} = \{c_i\}_{i=1}^{M}$ over a user set $\mathcal{U} = \{u_i\}_{i=1}^{N}$. Each cascade $c \in \mathcal{C}$ chronologically records the diffusion of a piece of information: $c = \{(u_1, t_1), (u_2, t_2), ..., (u_{|c|}, t_{|c|})\}$, where $t_{i-1} \leq t_i$. The tuple $(u_i, t_i)$ indicates that the user $u_i$ reposted the information at timestamp $t_i$.



Given a user set $\mathcal{U}$, we define the social graph as $\mathcal{G}_S = (\mathcal{U}, \mathcal{E})$, where $\mathcal{E}$ denotes social relations (e.g., follower/followee links). For a given cascade set $\mathcal{C}$, we construct a sequence of diffusion hypergraphs $\mathcal{G}_D = \{\mathcal{G}^t_D\}_{t=1}^T$ by partitioning $\mathcal{C}$ into $T$ subsets based on timestamps. Each hypergraph $\mathcal{G}^t_D = (\mathcal{U}^t, \mathcal{E}^t)$ records the diffusion activity within the $t$-th time interval, where $\mathcal{U}^t \subseteq \mathcal{U}$ denotes the users involved, and $\mathcal{E}^t$ is the set of hyperedges, with each hyperedge linking users who participated in the same cascade during that interval.

Based on the above introduction, the IDP tasks we focus on can be defined as follows:

\textbf{Definition 1.} Macroscopic Prediction - 
\textit{Given a social graph $\mathcal{G}_S$, temporal diffusion hypergraphs $\mathcal{G}_D$, and a snapshot of an information cascade $c$ observed at time $t_o$, the goal is to predict its final size $|c|$, also referred to as the cascade's popularity.}

\textbf{Definition 2.} Microscopic Prediction - \textit{Given the same input, the goal is to predict, at the next time step, which user is most likely to repost the content.}

\vspace{-0.4em}


\subsection{Vanilla Test-Time Training}
Test-Time Training, proposed by Sun et al.~\cite{sun2020test}, is a paradigm for addressing distribution shifts by adapting a trained model to each test instance. It uses a Y-shaped architecture with a shared feature extractor and two branches: one for the primary task and another for a self-supervised auxiliary task. TTT consists of three phases: joint training, test-time training, and inference.

\textbf{Joint training:} The model is trained via multi-task learning on a labeled training dataset, using a joint loss from the primary and auxiliary tasks to update all parameters. After training, we denote the shared feature extractor, the primary task head, and the auxiliary task head as $\theta_e$, $\theta_m$, and $\theta_s$, respectively.

\textbf{Test-Time Training:} Given an unlabeled test sample $x$, we first adapt the jointly trained feature extractor $\theta_e$ to $x$ by minimizing the auxiliary loss, while keeping the primary task head $\theta_m$ fixed:
\begin{equation}
\min_{\theta_e} \mathcal{L}_s(x; \theta_e, \theta_s).
\end{equation}
Let $\hat{\theta}_e$ denote the adapted feature extractor.

\textbf{Inference:} The final prediction for the primary task is made using $\hat{\theta}_e$ and $\theta_m$, expressed as $\hat{\theta}_e \circ \theta_m(x)$.



\section{METHODOLOGY}

In this section, we present our model \textit{T3MAL}, the first TTT-based framework designed for robust multi-scale information diffusion prediction under distribution shifts. Our goal is to enable instance-specific model adaptation at test time without requiring any manual labels. As shown in Fig.~\ref{fig:model_architecture}, \textit{T3MAL} consists of two modules: (a) \textit{user representation learning module}, and (b) \textit{primary and auxiliary task learning module}. The former learns user representations by encoding both the social graph and temporal diffusion hypergraphs. The latter includes a primary network for multi-scale information diffusion prediction and a BYOL-inspired auxiliary network for self-supervised learning. Both networks share a common feature extraction backbone, enabling the model to adapt to each test instance through the auxiliary loss. 

We first describe the details of each module in Sections~\ref{subsec:user representation learning} and~\ref{subsec:primary and auxiliary task learning}. Then, in Section~\ref{subsec:test-time fast adaptation}, we present the full training and testing pipeline, which consists of three phases: joint training, meta-auxiliary training, and meta-auxiliary testing. This pipeline highlights how meta-auxiliary learning enables fast and accurate test-time adaptation.

\vspace{-0.4em}
\subsection{User Representation Learning}
\label{subsec:user representation learning}

As shown in Fig. \ref{fig:model_architecture}(a), this module employs two encoders to separately model the social graph and temporal diffusion hypergraphs, yielding user representations with distinct structural semantics. Specifically, we adopt a multi-layer Graph Convolutional Network (GCN) \cite{kipf2016semi} and a Hypergraph Neural
Network (HGNN) \cite{cheng2023enhancing} as backbones. The GCN encodes relatively stable social relationships among users, while the HGNN captures global user interactions and diffusion dynamics.

\subsubsection{\textbf{Social Graph Encoder}} 
Given the social graph $\mathcal{G}_S = (\mathcal{U}, \mathcal{E})$, we initialize all user representations $\mathbf{X}_S^{0} \in \mathbb{R}^{N \times d}$ by sampling from a normal distribution, where $N$ denotes the number of users in $\mathcal{U}$. By stacking multiple layers of GCN, we obtain the social representations of the users, denoted as $\mathbf{X}_S \in \mathbb{R}^{N \times d}$. This social graph encoding effectively mitigates the \textit{cold-start} problem in prediction tasks by allowing the model to infer user preferences from social neighborhood
structures, even for users with sparse interaction history. 

\subsubsection{\textbf{Duffusion Hypergraph Encoder}} Given the temporal diffusion hypergraphs $\mathcal{G}_D$, we employ HGNN to encode each hypergraph $\mathcal{G}_D^t$ at time interval $t$, capturing global user preferences. The message passing involves two steps: node-to-hyperedge aggregation and hyperedge-to-node aggregation. First, we compute the representation $\mathbf{m}_{j,t}$ of hyperedge $e^t_j$ by aggregating the representations of its connected users. 
Then, since a user may participate in multiple cascades during the same interval, we update the representation $\mathbf{x}_{i,t}$ of user $u_i$ by aggregating messages from its associated hyperedges $\mathcal{E}_i^t$, where $\mathcal{E}_i^t = \{e_j^t \mid u_i \in e_j^t\}$ denotes the set of hyperedges that include user $u_i$ during time interval $t$. By stacking multiple HGNN layers, the model can capture higher-order user-cascade interactions, formalized as:


\begin{equation}
\begin{aligned}
\mathbf{m}_{j,t}^{l+1} &= \sigma \Big( \frac{1}{\left| e^t_j \right|}\sum_{u_i^t \in e_j^t} \mathbf{W}_u \cdot \mathbf{x}_{i,t}^l \Big),\\
\mathbf{x}_{i,t}^{l+1} &= \sigma \Big( \frac{1}{|\mathcal{E}^t_i|} \sum_{e_j^t \in \mathcal{E}^t_i} \mathbf{W}_e \cdot \mathbf{m}_{j,t}^{l+1} \Big),
\end{aligned}
\end{equation}
where $\mathbf{W}_u$ and $\mathbf{W}_e$ are the learnable weight matrices.

\begin{figure*}[t]
\centering
\includegraphics[trim=13 8 8 8, clip, width=\textwidth]{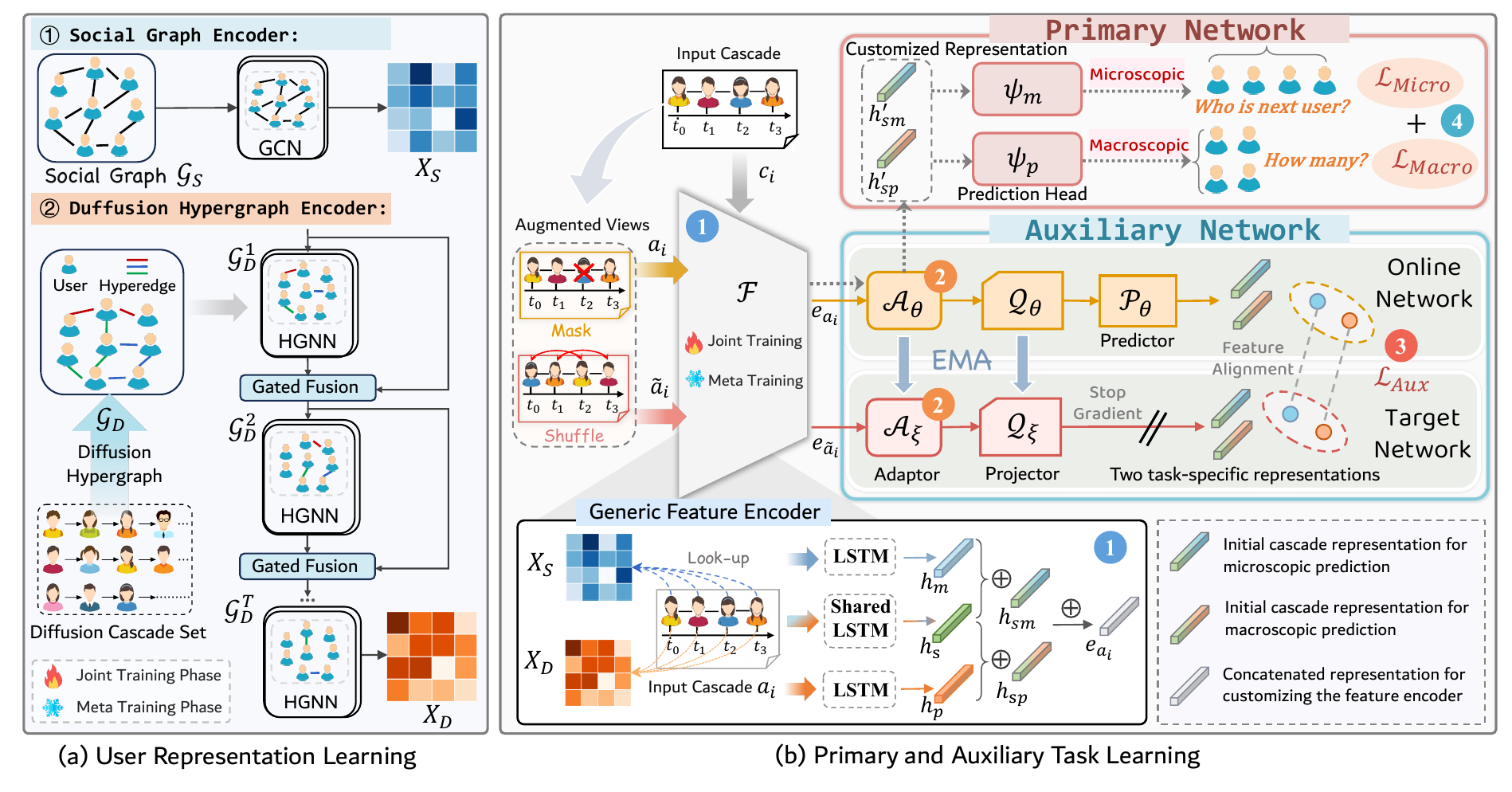}
\vspace{-1.3em} 
\caption{\textbf{Overview of the proposed T3MAL framework}. It comprises two modules: (a) \textit{User Representation Learning Module}, which encodes the social graph and temporal diffusion hypergraphs using GCN and HGNN to obtain two types of user representations. (b) \textit{Primary and Auxiliary Task Learning Module}, which includes a primary network for multi-scale diffusion prediction and a BYOL-inspired auxiliary network for self-supervised learning. Both share a common feature extraction backbone, including the generic feature encoder $\mathcal{F}$ and the adaptor $\mathcal{A}_{\theta}$, but have distinct prediction heads. \textcolor{customorange}{Orange} and \textcolor{customred}{red} arrows indicate the auxiliary network's processing of augmented views, while \textcolor{gray}{gray} dashed arrows show the primary network’s processing of the original input cascade.}

\label{fig:model_architecture}
\vspace{-2mm}  
\end{figure*}

A single $L$-layer HGNN only models cascade diffusion within a specific time interval. To capture the temporal evolution of cascades and user preference shifts, we introduce a gated fusion strategy to bridge user representations across adjacent intervals. 
Specifically, for each user $u_i$,  we fuse the initial embedding $\mathbf{x}^0_{i,t}$ with the updated embedding $\mathbf{x}^{L}_{i,t}$—learned from the current hypergraph $\mathcal{G}^t_D$—using learnable attention weights. The resulting fused embedding is then used as the initial state $\mathbf{x}^0_{i,t+1}$ for the next time interval. 
After all time intervals are processed, we obtain the final diffusion-aware user representations $\mathbf{X}_D \in \mathbb{R}^{N \times d}$.

\vspace{-0.4em}
\subsection{Primary and Auxiliary Task Learning}
\label{subsec:primary and auxiliary task learning}

Following the TTT paradigm, this module incorporates a primary network for diffusion prediction and an auxiliary network for self-supervised learning, as shown in Fig. \ref{fig:model_architecture}(b). Both networks share a common feature extraction backbone, which consists of a generic feature encoder and a adaptor. We detail the architectures and training objectives of both networks below.

\subsubsection{\textbf{Auxiliary Task Learning}} 

Learning primary tasks alongside a well-designed auxiliary task enables the model to extract richer, more informative features that provide complementary signals for the primary tasks. 
Given the supervision-free nature of TTT, a self-supervised auxiliary network is essential for guiding the model to learn features relevant to diffusion prediction without relying on manual labels. Here, we adopt a modified version of \textbf{BYOL} \cite{grill2020bootstrap} as the auxiliary objective. Unlike contrastive learning, BYOL avoids the need for negative sampling, making it particularly well-suited for TTT where only a single test cascade is available.



The BYOL architecture includes an online network and a target network, each processing different augmented views of the same cascade. The online network is trained to align its representation with that of the target network by minimizing the mean squared error between their $\ell_2$-normalized outputs. In our design, the online network (parameterized by $\theta$) comprises four components: a generic feature encoder $\mathcal{F}$, an adaptor $\mathcal{A}_\theta$, a projector $\mathcal{Q}_\theta$, and a predictor $\mathcal{P}_\theta$. The target network (parameterized by $\xi$) mirrors this structure but omits the predictor (see Fig. \ref{fig:model_architecture}(b)). It serves as a stable regression target and is updated via an exponential moving average (EMA) of $\theta$, rather than gradient-based optimization.


Unlike the traditional BYOL model \cite{grill2020bootstrap}, we treat the feature encoder $\mathcal{F}$ as a generic backbone shared by both networks and freeze it after joint training to preserve its encoded prior knowledge. While the trained $\mathcal{F}$ captures common diffusion patterns across cascades, each cascade may still exhibit unique traits. To address this, we introduce a lightweight adaptor to customize $\mathcal{F}$ for each input—whether original or augmented—enabling the model to better capture subtle, input-specific variations. This design not only aligns with the TTT paradigm but also helps mitigate catastrophic forgetting.

For IDP tasks, we design the feature encoder $\mathcal{F}$ with a shared LSTM to capture features common to both microscopic and macroscopic predictions, along with two task-specific LSTMs to learn features unique to each task. The adaptor, projector, and predictor are implemented as multi-layer perceptrons (MLPs). We choose LSTMs for their effectiveness in our framework and their lower complexity compared to transformers.

In the auxiliary network, we first generate two augmented views, $a_i$ and $\tilde{a}_i$, for each input cascade $c_i \in \mathcal{C}$ using sequence augmentation strategies (e.g., user masking and shuffling \cite{xie2022contrastive}). The view $a_i$ is fed into the online network, while $\tilde{a}_i$ is passed to the target network. Below, we take $a_i$ as an example to illustrate the operations of each component in the online network; the target network follows an analogous procedure.

\textit{- Generic Feature Encoder.}
For each user in the augmented view $a_i$, we first retrieve their corresponding social and diffusion-aware representations from $\mathbf{X}_S$ and $\mathbf{X}_D$, respectively, and arrange them in the order they appear in $a_i$. This yields two input sequences: $\mathbf{u}_S = [(\mathbf{x}_j)] \in \mathbb{R}^{|a_i| \times d}$ and $\mathbf{u}_D  = [(\mathbf{x}_j)] \in \mathbb{R}^{|a_i| \times d}$, which are then fed into the generic encoder $\mathcal{F}$ to capture diffusion dependencies among users for downstream prediction tasks. 
Specifically, at each time step of the shared LSTM, we take both the social and diffusion-aware representations of the current user as input to learn a shared representation $\mathbf{h}_{s} \in \mathbb{R}^{|a_i| \times d}$, which captures features common to both prediction tasks. In parallel, we employ two task-specific LSTMs to independently encode $\mathbf{u}_S$ and $\mathbf{u}_D$, yielding task-specific representations: $\mathbf{h}_m$ for microscopic prediction and $\mathbf{h}_p$ for macroscopic prediction.
Finally, we obtain the initial cascade representations by concatenating the shared and task-specific representations: $\mathbf{h}_{sm} = [\mathbf{h}_s \| \mathbf{h}_m]$ for microscopic prediction and $\mathbf{h}_{sp} = [\mathbf{h}_s \| \mathbf{h}_p]$ for macroscopic prediction.


\textit{- Adaptor.} 
The lightweight adaptor $\mathcal{A}_\theta$ is introduced to customize $\mathcal{F}$ for each input, whether it is an original cascade or an augmented view. This design helps the model preserve the common diffusion patterns learned during training while adapting to the unique characteristics of individual inputs.
Specifically, given the initial cascade representations $\mathbf{h}_{sm}$ and $\mathbf{h}_{sp}$ for augmentation $a_i$, we first concatenate them into a unified embedding $\mathbf{e}_{a_i}$, which serves as the input to the adaptor. The adaptor then generates a set of adaptation parameters $\Pi_{a_i}$ tailored for the three LSTMs in $\mathcal{F}$.

Let the original weight matrices of the three LSTMs in $\mathcal{F}$ be denoted by $\mathcal{F} = \{\mathbf{W}_s, \mathbf{W}_m, \mathbf{W}_p\}$, where each $\mathbf{W}_* \in \mathbb{R}^{d \times d}$. The corresponding adaptation parameters are given by $\Pi_{a_i} = \{\mathbf{\Pi}_s, \mathbf{\Pi}_m, \mathbf{\Pi}_p\}$, with each $\mathbf{\Pi}_* \in \mathbb{R}^{2d \times d}$. As an illustrative example, consider the adaptation of the shared LSTM. The updated weight matrix $\mathbf{\hat{W}}_s$ is computed as:
\begin{equation}
\mathbf{\hat{W}}_s = \mathbf{W}_s \diamond \mathbf{\Pi}_s,
\end{equation}
where $\diamond$ represents the adaptation operation. We adopt FiLM \cite{perez2018film} as the adaptation mechanism, where $\mathbf{\Pi}_s$ is split and reshaped into scaling and shifting parameters $\bm{\gamma}_s, \bm{\beta}_s \in \mathbb{R}^{d \times d}$, yielding:
\begin{equation}
\mathbf{\hat{W}}_s = \mathbf{W}_s \odot \bm{\gamma}_s + \bm{\beta}_s,
\end{equation}
where $\odot$ denotes element-wise multiplication. This process produces a customized feature encoder for $a_i$, denoted by $\hat{\mathcal{F}}_{a_i} = \{\mathbf{\hat{W}}_s, \mathbf{\hat{W}}_m, \mathbf{\hat{W}}_p\}$, and expressed as:
\begin{equation}
\hat{\mathcal{F}}_{a_i} = \mathcal{F} \diamond \Pi_{a_i}.
\end{equation}
Following the same procedure, a customized encoder $\hat{\mathcal{F}}_{\tilde{a}_i}$ is also derived for the alternative view $\tilde{a}_i$. 

\textit{- Projector and Predictor.} Using the customized feature
encoders $\hat{\mathcal{F}}_{a_i}$, we re-encode the input $a_i$ to obtain its customized cascade representations $\mathbf{h}_{sm}'$ and $\mathbf{h}_{sp}'$. These representations are passed through the projector $\mathcal{Q}_{\theta}$ to map them into a lower-dimensional space for feature alignment. The predictor $\mathcal{P}_{\theta}$ then further transforms the projected representations to better align with those of the target network, ensuring the consistency of feature representations across augmented views.


\textit{Auxiliary Task Objective.}
Following the steps above, we compute the online network's prediction for $a_i$ as $\mathbf{r}_{\theta} = \left[ \mathcal{Q}_{\theta} \circ \mathcal{P}_{\theta}(\mathbf{h}_{sm}'), \mathcal{Q}_{\theta} \circ \mathcal{P}_{\theta}(\mathbf{h}_{sp}') \right]$
 and the target network's projection for $\tilde{a}_i$ as $\mathbf{z}_{\xi} = \left[ \mathcal{Q}_{\xi}(\mathbf{h}_{sm}'), \mathcal{Q}_{\xi}(\mathbf{h}_{sp}') \right]$. Here, $\circ$ denotes operator composition. The objective is to minimize the mean squared error between the $\ell_2$-normalized online prediction $\mathbf{r}_{\theta}$ and the target projection $\mathbf{z}_{\xi}$, formulated as:
\begin{equation}
\mathcal{L}_{\theta, \xi} = 2 - \frac{2 \mathbf{r}_{\theta}^\top \mathbf{z}_{\xi}}{\| \mathbf{r}_{\theta} \|_2 \cdot \| \mathbf{z}_{\xi} \|_2}.
\end{equation}
Moreover,  we further define a symmetric loss term $\mathcal{\tilde{L}}_{\theta, \xi}$ by swapping the inputs of the online and target networks. Therefore, the final auxiliary network objective is:
\begin{equation}
\mathcal{L}_\text{Aux} = \mathcal{L}_{\theta, \xi} + \mathcal{\tilde{L}}_{\theta, \xi}.
\end{equation}

\subsubsection{\textbf{Primary Task Learning}} 
The primary network performs multi-scale diffusion predictions through two branches: one for macroscopic and one for microscopic prediction. Both branches share the feature extraction backbone with the auxiliary network, including the generic feature encoder $\mathcal{F}$ and adaptor $\mathcal{A}_{\theta}$. Each branch has its own prediction head for its respective task. Unlike the auxiliary network, the primary network takes the original cascade as input rather than an augmented view.

\textit{- macroscopic prediction branch.} Given an input cascade $c_i \in \mathcal{C}$, this branch aims to estimate its final popularity. To do so, we first compute its customized cascade representation $\mathbf{h}_{sp}' = \mathcal{F} \circ \mathcal{A}_\theta (c_i)$, which is then passed to a prediction head $\psi_p$ to generate the estimated popularity $\hat{y}_i$. Following prior works \cite{xu2021casflow, lu2023continuous}, we employ the Mean Squared Logarithmic Error (MSLE) loss to penalize the deviation between the predicted and actual popularity:
\begin{equation}
\mathcal{L}_\text{Macro} = \frac{1}{M} \sum_{i=1}^{M} \left( \log(1 + \hat{y}_i) - \log(1 + y_i) \right)^2,
\end{equation}
where $M$ denotes the number of cascades in $\mathcal{C}$.


\textit{- microscopic prediction branch.} The branch is similar to the macroscopic one but differs in its learning objective. It aims to predict the next user likely to engage with the cascade using a classification head $\psi_m$. This task is optimized by minimizing the cross-entropy loss:
\begin{equation}
\mathcal{L}_\text{Micro} = - \sum_{i=2}^{|c_i|} \sum_{j=1}^{N} y_{ij} \log(\hat{y}_{ij}),
\end{equation}
where $y_{ij}$ is a binary indicator of whether user $j$ is the $i$-th user to forward the cascade, and $\hat{y}_{ij}$ denotes the predicted probability that user $j$ appears at position $i$.

The overall loss of the primary network is defined as a weighted sum of the microscopic and macroscopic losses, with $\lambda \in [0, 1]$ controlling their trade-off:
\begin{equation}
\mathcal{L}_\text{Pri} = (1-\lambda )\mathcal{L}_\text{Micro} + \lambda \mathcal{L}_\text{Macro}.
\label{eq: primary task loss}
\end{equation}


\begin{figure*}[t]
    \centering

    \begin{subfigure}[b]{0.487\textwidth}
        \includegraphics[trim=1.5 0 1 4, clip, width=\linewidth]{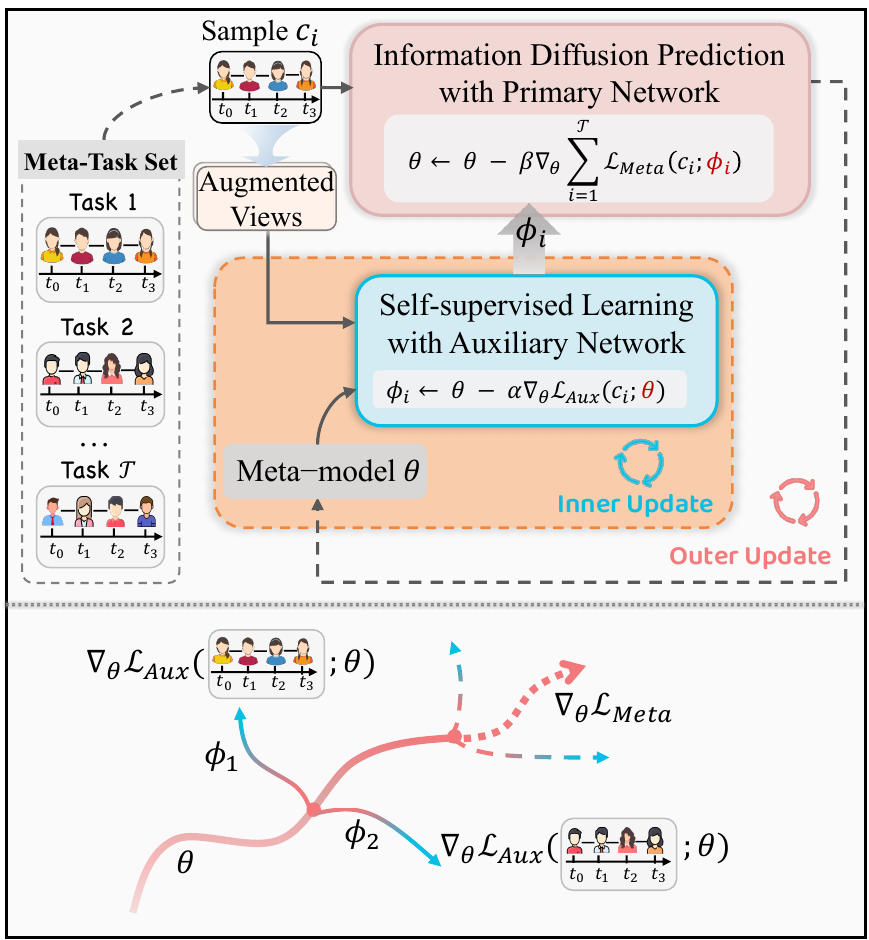}
        \caption{\small Meta-Auxiliary Training.}
        \label{fig: Meta-auxiliary training}
    \end{subfigure}
    \hfill
    \begin{subfigure}[b]{0.487\textwidth}
        \includegraphics[trim=1 0 1 0.8, clip, width=\linewidth]{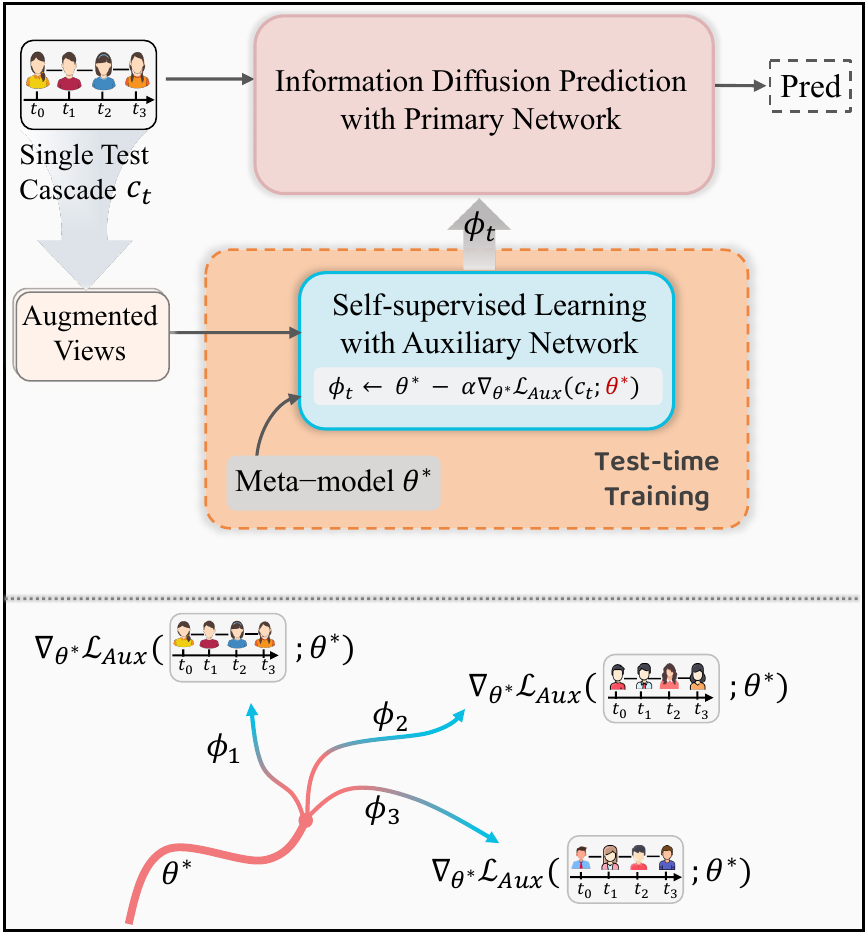}
        \caption{\small Meta-Auxiliary Testing.}
         \label{fig: Meta-auxiliary testing}
    \end{subfigure}

    \caption{
        \textbf{Overview of the proposed Meta-Auxiliary Learning scheme}. (a) \textit{Meta-Auxiliary Training Phase}: For each training cascade $c_i$, the meta-model $\theta$ is first adapted via the auxiliary loss in the inner loop, yielding $\phi_i$. Next, $\phi_i$ is evaluated on the primary tasks. Finally, we update the meta-model $\theta$ in the outer loop by minimizing the joint losses across all $\mathcal{T}$ tasks. (b) \textit{Meta-Auxiliary Testing Phase}: For each test cascade, we use the auxiliary network to fine-tune the well-trained meta-model $\theta^*$ and then use the adapted model for prediction.
    }
    \label{fig:Meta-Auxiliary Learning}
    \vspace{-2mm}  
\end{figure*}

\vspace{-0.5em}
\subsection{Test-Time Fast Adaptation}
\label{subsec:test-time fast adaptation}
To enable efficient model adaptation at test time, we introduce a meta-auxiliary learning scheme that explicitly correlates the performance of the primary and auxiliary tasks to facilitate optimal weight initialization. This approach overcomes the limitations of vanilla TTT methods and enables fast, accurate adaptation of a trained model to unseen test cascades. The optimization consists of three phases: joint training, meta-auxiliary training, and meta-auxiliary testing, as detailed below.


\subsubsection{\textbf{Joint Training}} 
It is quite challenging to perform meta-auxiliary training from scratch. Therefore, we first jointly train the primary and auxiliary tasks on the training set $\mathcal{D}^{tr}$ to obtain a jointly trained model. The loss is defined as a linear combination of the primary and auxiliary losses with hyperparameter $\gamma$:
\begin{equation}
\mathcal{L}_\text{Joint} = \mathcal{L}_\text{Pri} + \gamma \mathcal{L}_\text{Aux}.
\label{eq:joint_training loss}
\end{equation}

At this stage, all parameters are updated via gradient descent, while the target network parameters $\xi$ (including the adaptor $\mathcal{A}_\xi$ and projector $\mathcal{Q}_\xi$) are updated using an exponential moving average of the online network parameters $\theta$:
\begin{equation}
\xi \leftarrow \tau \xi + (1 - \tau) \theta,
\end{equation}
where $\tau$ is the target decay rate.

The jointly trained model, obtained from Eq.\ \eqref{eq:joint_training loss}, serves as the initialization for meta-auxiliary learning.  

\subsubsection{\textbf{Meta-auxiliary Training}} \label{sec:meta_auxiliary training}

Our empirical analysis in Section~\ref{subsec:complexity and efficiency analysis} reveals a key limitation of the vanilla TTT paradigm: the jointly trained model optimized with $\mathcal{L}_\text{Joint}$ may not be well-suited for adapting to specific test cascades. This limitation likely arises from the model’s lack of foresight into future learning objectives, such as adapting to unseen test cascades. Consequently, when distribution shifts occur at test time, TTT may require substantial and unpredictable gradient updates. Furthermore, the updates driven by the auxiliary task do not consistently benefit the primary tasks, and the model may become biased towards optimizing the auxiliary task at the expense of the primary objectives. 



To address this, we propose \textit{meta-auxiliary training}, which aims to learn an optimal weight initialization for TTT, enabling fast adaptation to unseen test cascades with only a few gradient updates. For simplicity, let $\theta = \{ \mathcal{F}, \mathcal{A}_\theta, \mathcal{Q}_\theta, \mathcal{P}_\theta, \psi_p,\psi_m \}$ denote the meta-model, initialized from the jointly trained model. Here, $\mathcal{F}$ and $\mathcal{A}_\theta$ form the shared backbone between the primary and auxiliary networks, while $\psi_p$ and $\psi_m$ are the prediction heads for the two primary tasks. To reduce overhead and retain prior knowledge, we freeze the user representation learning module and the encoder $\mathcal{F}$, updating only the rest of $\theta$ during meta-auxiliary training.

During each iteration of meta-auxiliary training, a batch of $\mathcal{T}$ cascades is sampled from the training set $\mathcal{D}^{{tr}}$, with each cascade $c_i$ treated as an individual meta-task. As shown in Fig.~\ref{fig: Meta-auxiliary training}, the procedure follows a nested-loop structure. In the inner loop, the meta-model $\theta$ is adapted to each task by minimizing the auxiliary loss, yielding a task-specific model $\phi_i$ after a few gradient updates. In the outer loop, these adapted models are evaluated using a joint loss that combines both primary and auxiliary objectives. The meta-model $\theta$ is then updated by aggregating the joint losses across all $\mathcal{T}$ tasks. This design explicitly couples the auxiliary objective with IDP task performance, ensuring that parameter updates driven by the auxiliary loss also benefit the primary tasks, thereby mitigating the drawbacks of vanilla TTT.

Concretely, in the inner loop, we first compute the self-supervised auxiliary loss $\mathcal{L}_\text{Aux}(c_i;\theta)$ for each input cascade $c_i$ in the batch. In this context, both the online and target networks are initialized from the meta-model $\theta$. The task-specific model $\phi_i$ acts as the online network that requires adaptation to cascade $c_i$. Unlike the BYOL setup described previously, the meta-model $\theta$—as a smoothed version of the task-specific models $\{\phi_i\}_i^{\mathcal{T}}$—can directly serve as the target network. As a result, a separate target network $\xi$ is no longer necessary and the target network parameters are not updated via EMA.
For the two augmented views $a_i$ and $\tilde{a}_i$, we denote the online predictions from $\phi_i$ as $\mathbf{r}_{\phi_i}^i$ and $\tilde{\mathbf{r}}_{\phi_i}^i$, and the corresponding target projections from $\theta$ as $\mathbf{z}_{\theta}^i$ and $\tilde{\mathbf{z}}_{\theta}^i$. The auxiliary loss for optimizing $\phi_i$ is defined as:
\begin{equation}
\mathcal{L}_\text{Aux}(c_i;\theta) = \mathcal{L}_{\phi_i, \theta}^{i} + \mathcal{\tilde{L}}_{\phi_i, \theta}^{i},
\end{equation}
\begin{equation}
\text{with} \quad \mathcal{L}_{\phi_i, \theta}^{i} = 2 - 2 \cdot \frac{(r_{\phi_i}^i)^T \tilde{z}_{\theta}^i}{\|r_{\phi_i}^i\|_2 \cdot \|\tilde{z}_{\theta}^i\|_2}.
\label{eq: auxiliary loss}
\end{equation}

For each meta-task, we adapt the meta-model $\theta$ via $\delta$ steps of gradient descent on the auxiliary loss, yielding a task-specific model $\phi_i$ as follows:
\begin{equation}
    \phi_i \leftarrow \theta - \alpha \nabla_{\theta} \mathcal{L}_\text{Aux}(c_i;\theta),
\label{eq: innner update loss}
\end{equation}
where $\alpha$ denotes the inner learning rate.

Let the adapted task-specific model be denoted as $\phi_i = \{\mathcal{F}, \mathcal{A}_{\phi_i}, \mathcal{Q}_{\phi_i}, \mathcal{P}_{\phi_i}, \psi_p,\psi_m \}$, where only $\{\mathcal{A}_{\phi_i}, \mathcal{Q}_{\phi_i}, \mathcal{P}_{\phi_i}\}$ are updated via Eq.\ \eqref{eq: innner update loss}. 
The adapted model $\phi_i$, optimized through the auxiliary task, is also expected to improve performance on the primary tasks. Therefore, in the outer loop, we aim to evaluate $\phi_i$ based on the primary task loss $\mathcal{L}_{pri}(c_i;\phi_i)$. However, since the primary tasks do not directly involve the projector $\mathcal{Q}_{\theta}$ and the predictor $\mathcal{P}_{\theta}$, the primary task loss alone does not provide the gradient signals necessary to update these modules. Thus, we instead evaluate $\phi_i$ using a joint loss that incorporates both the auxiliary and primary task objectives. The meta-objective for optimizing the meta-model $\theta$ is defined as:
\begin{equation}
\mathcal{L}_\text{Meta} = \mathcal{L}_\text{Pri}(c_i;\phi_i) + \gamma \cdot \mathcal{L}_\text{Aux}(c_i;\phi_i),
\label{eq:meta-model loss}
\end{equation}
\begin{equation}
    \theta \leftarrow \theta - \beta \nabla_\theta \sum_{i=1}^{\mathcal{T}} \mathcal{L}_\text{Meta}(c_i;\phi_i),
\end{equation}
where $\beta$ represents the meta-learning rate in the outer loop. 
The full meta-auxiliary training procedure is illustrated in Fig.~\ref{fig: Meta-auxiliary training} and summarized in \textit{\textbf{Appendix E}}.

After completing meta-auxiliary training, we obtain the meta-trained model $\theta^*$.

\subsubsection{\textbf{Meta-auxiliary Testing}} 

At this stage, we perform test-time training on each test cascade $c_t$. Specifically, we adapt the meta-trained model $\theta^*$ to $c_t$ by minimizing the auxiliary loss (Eq.\ \eqref{eq: innner update loss}), and use the adapted model $\phi_t$ to make predictions for both primary tasks. Afterward, $\phi_t$ is discarded, and the model is reset to $\theta^*$ before adapting to the next test cascade.

\section{Experiments}

\vspace{0.3em}
\subsection{Experimental Settings}  

\noindent{\textbf{Datasets}}. Following prior research \cite{feng2022h,qiao2023rotdiff,sun2022ms}, we conducted experiments on three publicly available datasets: Android \cite{sankar2020inf}, Twitter \cite{hodas2014simple}, and Douban \cite{zhong2012comsoc}, each with an underlying social graph. These datasets are collected from diverse sources: social media platforms (Twitter and Douban) and the Stack Exchange Q\&A website (for the Android dataset). For detailed descriptions of each dataset, please refer to \textit{\textbf{Appendix B}}. Table \ref{table:datasets_statistics} summarizes their key statistics: \# Users (total number of users), \# Links (number of connections in the social graph), \# Cascades (number of diffusion cascades), and Avg.Length (average cascade length). For each dataset, we chronologically split cascades into 80\% for training, 10\% for validation, and 10\% for testing.

\vspace{0.4em}
\noindent{\textbf{Baselines}}. To demonstrate T3MAL's effectiveness, we compare it against representative methods from three categories. For macroscopic prediction, the baselines include DeepCas \cite{li2017deepcas}, DeepHawkes \cite{cao2017deephawkes}, CasCN \cite{chen2019information}, and CasFlow \cite{xu2021casflow}. For microscopic prediction, we consider TopoLSTM \cite{wang2017topological}, NDM \cite{yang2019neural}, Inf-VAE \cite{sankar2020inf}, DyHGCN \cite{yuan2021dyhgcn}, MS-HGAT \cite{sun2022ms}, CE-GCN \cite{wang2022cascade}, and RotDiff \cite{qiao2023rotdiff}. For multi-scale prediction, we compare against FOREST \cite{yang2021full} and MINDS \cite{jiao2024enhancing}.

Moreover, to evaluate T3MAL’s robustness to distribution shifts in IDP tasks, we also compare it with the state-of-the-art (SOTA) UDA and DG methods, including DANN \cite{ganin2016domain}, MLDG \cite{li2018learning}, SFA \cite{li2021simple}, and CLUDA \cite{ozyurt2023contrastive}. 


For detailed descriptions of all baseline methods used in this study, please refer to the \textit{\textbf{Appendix C}}.

\noindent{\textbf{Evaluation Metrics}}. Following prior works \cite{yuan2021dyhgcn,xu2021casflow,qiao2023rotdiff}, we treat macroscopic prediction as a regression task and microscopic prediction as a retrieval task. For macroscopic prediction, we report the \textit{Mean Squared Logarithmic Error (MSLE)}. For microscopic prediction, we employ \textit{Hits@$k$} and \textit{MAP@$k$} with $k \in \{10, 50, 100\}$.

\begin{table}[t]
\caption{The statistics of three datasets.}
\centering
\resizebox{\linewidth}{!}{
\begin{tabular}{l|cccc}
\toprule
\textbf{Datasets} & \textbf{Android} & \textbf{Twitter} & \textbf{Douban} \\
\midrule
\# Users       & 9,958   & 12,627    & 12,232 \\
\# Links       & 48,573  & 309,631  & 198,496 \\
\# Cascades    & 679     & 3,442   & 3,475 \\
Avg.Length    & 33.3    & 32.60    & 21.76 \\
\bottomrule
\end{tabular}}
\label{table:datasets_statistics}
\vspace{-2mm}
\setlength{\belowcaptionskip}{-2mm}
\end{table}

\vspace{0.4em}


\begin{table*}[t]
    \centering
    \caption{Experimental results for microscopic prediction are reported using \textbf{\textit{Hits@k}}, where higher values indicate better performance. The best and second-best results are emphasized in \textbf{bold} and  \underline{underline}, respectively. Improvements over the top-performing baselines are statistically significant (sign test, $p < 0.01$).  
    }
    \label{tab:microscopic_results_hits}
    \resizebox{\linewidth}{!}{
    \begin{tabular}{l|ccc|ccc|ccc}
    \toprule
    \textbf{Models} & \multicolumn{3}{c|}{\textbf{Android}} & \multicolumn{3}{c|}{\textbf{Twitter}} & \multicolumn{3}{c}{\textbf{Douban}}\\
    \cmidrule{2-10}
    \textit{Hits@k} $\uparrow$ & \textbf{@10} & \textbf{@50} & \textbf{@100} & \textbf{@10} & \textbf{@50} & \textbf{@100} & \textbf{@10} & \textbf{@50} & \textbf{@100} \\
    \midrule
    TopoLSTM & 0.0460 & 0.1318 & 0.2103 & 0.0845 & 0.1580 & 0.2542 & 0.0857 & 0.1653 & 0.2147 \\
    NDM & 0.0170 & 0.0423 & 0.0555 & 0.1521 & 0.2823 & 0.3230 & 0.1000 & 0.2113 & 0.3014 \\
    Inf-VAE & 0.0318 & 0.0938 & 0.1452 & 0.1485 & 0.3272 & 0.4572 & 0.0894 & 0.2202 & 0.3572 \\
    DyHGCN & 0.0748 & 0.1746 & 0.2596 & 0.3188 & 0.4505 & 0.5219 & 0.1871 & 0.3233 & 0.3971 \\
    MS-HGAT & 0.1041 & 0.2031 & 0.2755 & 0.3350 & 0.4959 & 0.5891 & 0.2133 & 0.3525 & 0.4275 \\
    CE-GCN & 0.0886 & 0.1992 & 0.2727 & 0.3148 & 0.5087 & \underline{0.6117} & 0.1885 & 0.3272 & 0.4047 \\
    RotDiff & \underline{0.1138} & \underline{0.2301} & \underline{0.3126} & \underline{0.3582} & \underline{0.5239} & 0.6112 & \underline{0.2211} & \underline{0.3817} & \underline{0.4629} \\
    \midrule
    FOREST & 0.0866 & 0.1739 & 0.2314 & 0.2867 & 0.4207 & 0.4975 & 0.1950 & 0.3203 & 0.3908 \\
    MINDS & 0.1088 & 0.2090 & 0.2805 & 0.2972 & 0.4426 & 0.5321 & 0.1956 & 0.3087 & 0.3641 \\
    \midrule
    \rowcolor[HTML]{F2F2F2}
    \textbf{T3MAL} & \textbf{0.1449} & \textbf{0.2517} & \textbf{0.3372} & \textbf{0.3748} & \textbf{0.5420} & \textbf{0.6335} & \textbf{0.2585} & \textbf{0.4056} & \textbf{0.4953} \\
    \bottomrule
    \end{tabular}}
\end{table*}


\begin{table*}[t]
    \centering
    \caption{Experimental results for microscopic prediction are reported using \textbf{\textit{MAP@k}}, where higher values indicate better performance. The best and second-best results are emphasized in \textbf{bold} and  \underline{underline}, respectively. Improvements over the top-performing baselines are statistically significant (sign test, $p < 0.01$). }
    \label{tab:microscopic_results_map}
    \resizebox{\linewidth}{!}{
    \begin{tabular}{l|ccc|ccc|ccc}
    \toprule
    {\textbf{Models}} & \multicolumn{3}{c|}{\textbf{Android}} & \multicolumn{3}{c|}{\textbf{Twitter}} & \multicolumn{3}{c}{\textbf{Douban}} \\
    \cmidrule{2-10}
    \textit{MAP@k} $\uparrow$ & \textbf{@10} & \textbf{@50} & \textbf{@100} & \textbf{@10} & \textbf{@50} & \textbf{@100} & \textbf{@10} & \textbf{@50} & \textbf{@100}\\
    \midrule
    TopoLSTM & 0.0166 & 0.0202 & 0.0213 & 0.0851 & 0.1268 & 0.1368 & 0.0657 & 0.0753 & 0.0778 \\
    NDM & 0.0059 & 0.0070 & 0.0072 & 0.1241 & 0.1323 & 0.1430 & 0.0824 & 0.0873 & 0.0914 \\
    Inf-VAE & 0.0076 & 0.0103 & 0.0110 & 0.1980 & 0.2066 & 0.2132 & 0.1102 & 0.1128 & 0.1228 \\
    DyHGCN & 0.0392 & 0.0434 & 0.0446 & 0.2087 & 0.2148 & 0.2158 & 0.1061 & 0.1126 & 0.1136 \\
    MS-HGAT & 0.0639 & 0.0687 & 0.0696 & 0.1408 & 0.1504 & 0.1519 & \underline{0.1172} & \underline{0.1252} & \underline{0.1260} \\
    CE-GCN & 0.0477 & 0.0524 & 0.0534 & 0.1931 & 0.2020 & 0.2035 & 0.1103 & 0.1164 & 0.1175 \\
    RotDiff & \underline{0.0696} & \underline{0.0745} & \underline{0.0756} & \underline{0.2401} & \underline{0.2476} & \underline{0.2488} & 0.1164 & 0.1247 & 0.1258 \\
    \midrule
    FOREST & 0.0628 & 0.0667 & 0.0675 & 0.1960 & 0.2021 & 0.2175 & 0.1126 & 0.1184 & 0.1194 \\
    MINDS & 0.0680 & 0.0725 & 0.0735 & 0.1783 & 0.1849 & 0.1861 & 0.1142 & 0.1199 & 0.1213 \\
    \midrule
    \rowcolor[HTML]{F2F2F2}
    \textbf{T3MAL} & \textbf{0.0850} & \textbf{0.0873} & \textbf{0.0886} & \textbf{0.2545} & \textbf{0.2600} & \textbf{0.2617} & \textbf{0.1382} & \textbf{0.1449} & \textbf{0.1465} \\
    \bottomrule
    \end{tabular}}
\end{table*}

\vspace{0.4em}
\noindent{\textbf{Implementation Details.}}
We implement T3MAL in PyTorch and train it using the Adam optimizer, selecting optimal hyperparameters via grid search based on validation performance. The final results are reported on the best-validation checkpoint.
The maximum cascade length is set to 200, and the embedding dimension $d$ for both users and cascades is set to 64. We use a two-layer GCN to encode the social graph and a single-layer HGNN to model diffusion hypergraphs. 
For joint training, the batch size is set to 64, and the initial learning rate is $0.001$. The loss weights $\lambda$ (Eq.\eqref{eq: primary task loss}) and $\gamma$ (Eq.\eqref{eq:joint_training loss}) are set to 0.3 and 0.1, respectively.
For meta-auxiliary training, the meta-batch size $\mathcal{T}$ is set to 5, and the inner optimization is performed with two gradient steps ($\delta=2$). 
The learning rates for the inner and meta updates are set to $\alpha = 0.0005 $ and $\beta = 0.0002$ (Eq.\eqref{eq: innner update loss} and Eq.\eqref{eq:meta-model loss}).
All experiments are repeated five times, and we report the average results for accurate evaluation. For baselines, we adopt the default hyperparameters from their original papers and report the best results from either our reproduction or previously published papers.



\vspace{-0.5em}
\subsection{Main Results and Analysis} 
\textbf{Performance Comparison with SOTA.}
Evaluations on the microscopic (\Cref{tab:microscopic_results_hits,tab:microscopic_results_map}) and macroscopic (\Cref{tab:macroscopic_results}) prediction tasks yield three key observations:

\textbf{($\mathcal{O}1$)} As illustrated in \Cref{tab:microscopic_results_hits,tab:microscopic_results_map}, T3MAL consistently outperforms all state-of-the-art baselines on the microscopic prediction task, achieving relative improvements of at least 3.56\% in Hits@100 and 5.18\% in MAP@100 over the best-performing baselines. This superior performance can be primarily attributed to its multi-task learning framework, particularly the incorporation of the auxiliary task, which facilitates the learning of more robust and informative cascade representations. Moreover, models that jointly encode social relationships and cascade structures (e.g., T3MAL, MINDS, MS-HGAT) outperform those using only cascade sequences (e.g., NDM and TopoLSTM), highlighting the necessity of modeling both for effective microscopic prediction.

\begin{table}[t]
\caption{Results of macroscopic prediction are evaluated by MSLE, where lower values indicate better performance.}
\label{tab:macroscopic_results}
\centering
\resizebox{\linewidth}{!}{
\begin{tabular}{l|ccc}
\toprule
\textbf{Model ( \textit{MSLE}  $\downarrow$) }  & \textbf{Android} & \textbf{Twitter} & \textbf{Douban}\\
\midrule
DeepCas & 2.122 & 2.311 & 2.122 \\
DeepHawkes & 1.971 & 1.110 & 1.725 \\
CasCN & 0.981 & - & 1.476 \\
CasFlow & 1.041 & - & 0.465 \\
\midrule
FOREST & 0.556 & 0.847 & 0.825 \\
MINDS & \underline{0.151} & \underline{0.595} & \underline{0.404} \\
\midrule
\rowcolor[HTML]{F2F2F2}
\textbf{T3MAL} & \textbf{0.116} & \textbf{0.523} & \textbf{0.277} \\
\bottomrule
\end{tabular}
}
\label{table:models_comparison}
\end{table}

\textbf{($\mathcal{O}2$)} As reported in \Cref{tab:macroscopic_results}, T3MAL achieves relative MSLE reductions of 23.18\%, 12.10\%, and 31.44\% over the strongest baseline (MINDS) on three datasets, respectively. This can be attributed to two key factors: 1) the diffusion hypergraph encoder, which effectively captures global user behavior; and 2) the two-stage training strategy (joint training followed by meta-auxiliary learning), which facilitates full exploitation of all available dataset information and thus improves prediction.


\textbf{($\mathcal{O}3$)} Compared to other multi-scale models (e.g., FOREST, MINDS), T3MAL achieves better generalization performance on both prediction tasks by integrating TTT with meta-auxiliary learning.

\begin{figure}[t]
    \centering
    \begin{subfigure}[b]{0.485\linewidth}
        \includegraphics[width=\linewidth]{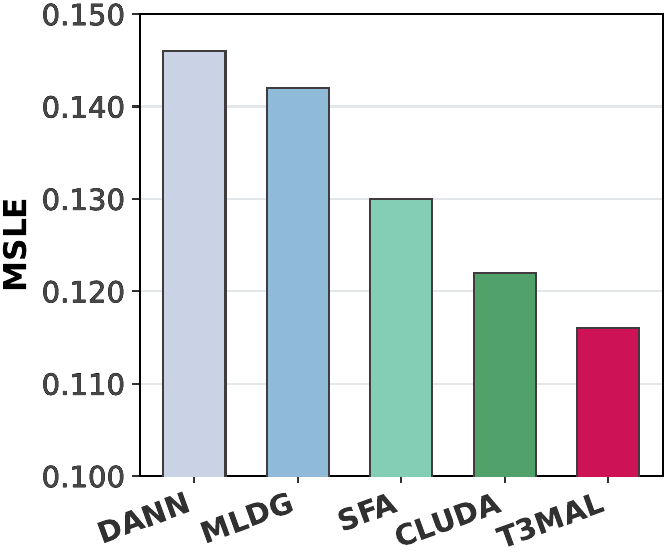}
        \caption*{\scriptsize (a) Android Dataset.}
    \end{subfigure}
    \hfill
    \begin{subfigure}[b]{0.485\linewidth}
        \includegraphics[width=\linewidth]{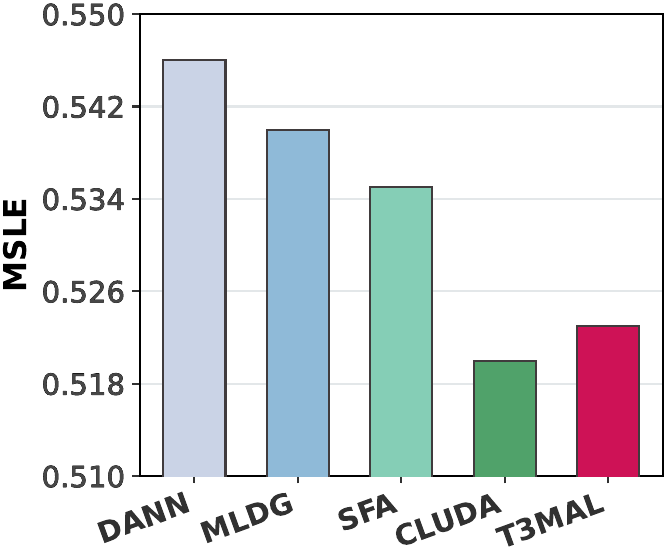}
        \caption*{\scriptsize (b) Twitter Dataset.}
    \end{subfigure}
    \caption{Comparison of macroscopic prediction performance among T3MAL, UDA (DANN, CLUDA), and DG (MLDG, SFA) methods.}
    \label{fig:generalization_comparison}
    \vspace{-2mm}  
\end{figure}

\textbf{Performance Comparison with UDA and DG.} 
To demonstrate the effectiveness of TTT in handling distribution shifts, we compare T3MAL with representative UDA and DG methods on the macroscopic prediction task using the Android and Twitter datasets. As shown in Figure~\ref{fig:generalization_comparison}, T3MAL is not only competitive but often outperforms these baselines. 
This performance advantage aligns with our motivation. Unlike UDA and DG methods, which address distribution shifts solely through complex model designs during training, T3MAL can flexibly adapt a trained model to newly encountered test cascades before making prediction. This enables better generalization to unseen distributions.

\vspace{-0.4em}
\subsection{Ablation Study}
To evaluate the contribution of each component in T3MAL, we conduct ablation studies on the Android and Twitter datasets, analyzing the effects of different task combinations and training strategies. The results for the macroscopic prediction tasks are presented in \Cref{table: macroscopic_ablation_study_results}, while the microscopic prediction results are provided in \textit{\textbf{Appendix F}} due to spatial constraints, where, once again, including all components yields the best performance.


\textbf{1) Choice of Auxiliary Task.}
The success of TTT hinges on selecting a suitable self-supervised auxiliary task that can extract useful features for the primary tasks with limited gradient updates. We compare three designs: (i) contrastive learning, which treats augmented views of the same cascade as positives and different cascades as negatives; (ii) MAE-based self-reconstruction \cite{zhu2025ghidorah}; and (iii) our BYOL variant. As detailed in \Cref{table: macroscopic_ablation_study_results} (rows 6 and 8), T3MAL achieves the best performance when using the BYOL variant as its auxiliary task.


\textbf{2) Is Multi-Task Learning Beneficial?} Ablation results in \Cref{table: macroscopic_ablation_study_results} confirm the effectiveness of our multi-task framework. Firstly, the BYOL-inspired auxiliary task enhances primary task performance: the “Macro + BYOL” variant achieves an average 3.54\% MSLE reduction over the “Macro” variant. Secondly, joint training of macroscopic and microscopic tasks enables synergistic optimization via cross-scale feature fusion, as evidenced by the comparison within the “Joint Training” type (see row 3 of \Cref{table: macroscopic_ablation_study_results}). These findings indicate that combining multi-scale prediction objectives with an auxiliary task yields richer feature representations, facilitating the modeling of complex diffusion patterns.



\begin{table}[t]
    \caption{Ablation study results on Android and Twitter datasets for macroscopic prediction task, evaluated by the MSLE metric.}
    \label{table: macroscopic_ablation_study_results}
    \begin{center}
    \begin{small}
    \resizebox{1.0\columnwidth}{!}{
    \begin{tabular}{p{2cm}lcc}
    \toprule
    \textbf{Type} & \textbf{Variant} & \textbf{Android} & \textbf{Twitter} \\
    \midrule
    \multirow{1}{*}{\centering Primary Only} & - Macro & 0.160 & 0.575 \\
    \midrule
    \multirow{2}{*}{\centering Joint Training} & - Macro + BYOL & 0.152 & 0.563 \\
    & - Macro + Micro + BYOL & 0.144 & 0.548 \\
    \midrule
    \multirow{2}{*}{\centering Meta-Training} & - Macro + BYOL + MT & 0.141 & 0.551 \\
    & - Macro + Micro + BYOL + MT & 0.131 & 0.537 \\
    \midrule
    \multirow{3}{*}{\centering Meta-Testing} & - Macro + BYOL + TTT & 0.142 & 0.577 \\
    & - Macro + Micro + BYOL + TTT & 0.130 & 0.565 \\
    & - Macro + BYOL + MT + TTT & 0.129 & 0.540 \\
    \midrule
    \multirow{2}{*}{\centering Auxiliary Task} & - Contrastive Learning & 0.132 & 0.543 \\
    & - MAE-based Self-Reconstruction & 0.125 & 0.534 \\
    \midrule
    \multirow{1}{*}{\centering BYOL Variant} & - w/o Adaptor & 0.122 & 0.530 \\
    \midrule
    \addlinespace
    \multirow{1}{*}{\centering \textbf{T3MAL}} & \textbf{All} & \textbf{0.116} & \textbf{0.523} \\
    \bottomrule
    \end{tabular}
    }
    \end{small}
    \end{center}
\end{table}




\textbf{3) Is Meta-Training Important?}
We compare three settings to assess the necessity of meta-auxiliary training: (i) joint training only, (ii) joint training with meta-auxiliary training but without TTT, and (iii) joint training with TTT but without meta-auxiliary training. 
As shown in~\Cref{table: macroscopic_ablation_study_results}, meta-trained models (i.e., those in the “Meta-Training” type) consistently outperform their jointly trained counterparts (i.e., those in the “Joint Training” type) by an average 5.10\% MSLE reduction, even without TTT. This indicates that meta-auxiliary training helps learn a better initialization, as described in Sec.~\ref{sec:meta_auxiliary training}. 
Moreover, directly applying TTT to jointly trained models without meta-training (i.e., the “Macro + BYOL + TTT” and “Macro + Micro + BYOL + TTT” variants in the “Meta-Testing” type) results in undesirable  performance degradation on the Twitter dataset, primarily due to catastrophic forgetting and overfitting to the auxiliary task (detailed analysis in Subsec.~\ref{subsec:analysis of catastrophic forgetting}).
These findings confirm that the meta-auxiliary training framework is essential not only for mitigating model bias towards the auxiliary task during TTT, but also for learning a good initialization for TTT.  

\textbf{4) Is TTT Effective?}
To validate the effectiveness of our TTT paradigm, we compare model variants with and without TTT. As shown in \Cref{table: macroscopic_ablation_study_results}, the observed performance gap (e.g., comparing “Macro + BYOL” against “Macro + BYOL + TTT”) indicates that TTT effectively improves model performance and enhances generalization (see Subsec.~\ref{subsec:visualization} for further analysis). 

\textbf{5) Is the Adaptor Effective?}  
To assess the impact of the adaptor, we ablate it from T3MAL and compare the results with the full model. As shown in the last two rows of \Cref{table: macroscopic_ablation_study_results}, excluding the adaptor leads to performance degradation, which highlights its effectiveness in customizing the feature encoder for each input cascade and mitigating catastrophic forgetting during TTT.

We also provide a detailed \textit{\textbf{hyperparameter sensitivity analysis}} in \textit{\textbf{Appendix G}}, which shows that our method is robust to a wide range of hyperparameter choices.


%




\begin{figure}[t]
    \centering
    \begin{subfigure}[b]{0.48\linewidth}
        \includegraphics[width=\linewidth]{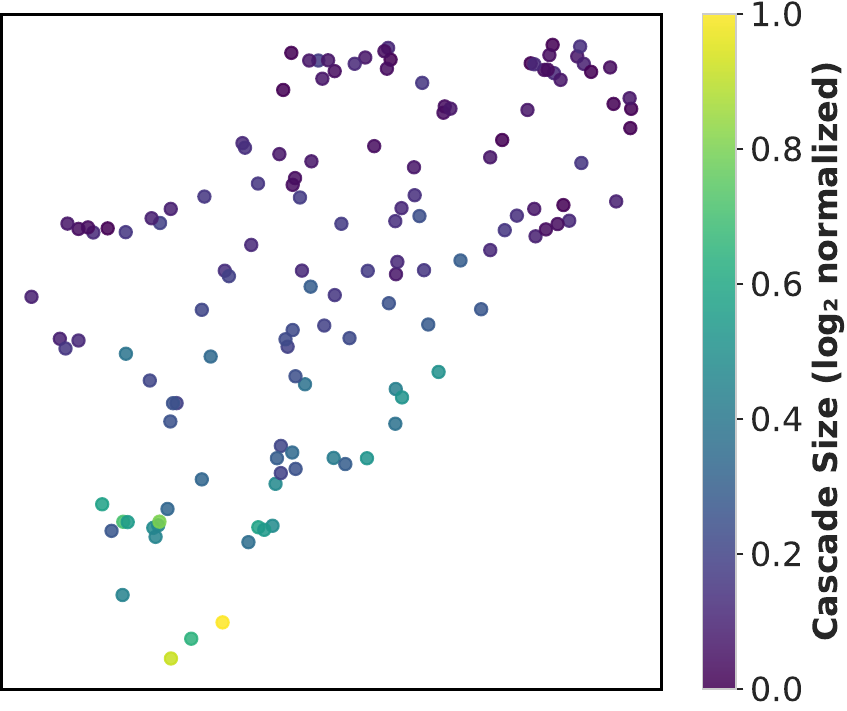}
        \caption*{\scriptsize (a) T3MAL w/o TTT.}
        \label{fig:macro_umap_before}
    \end{subfigure}
    \hfill
    \begin{subfigure}[b]{0.48\linewidth}
        \includegraphics[width=\linewidth]{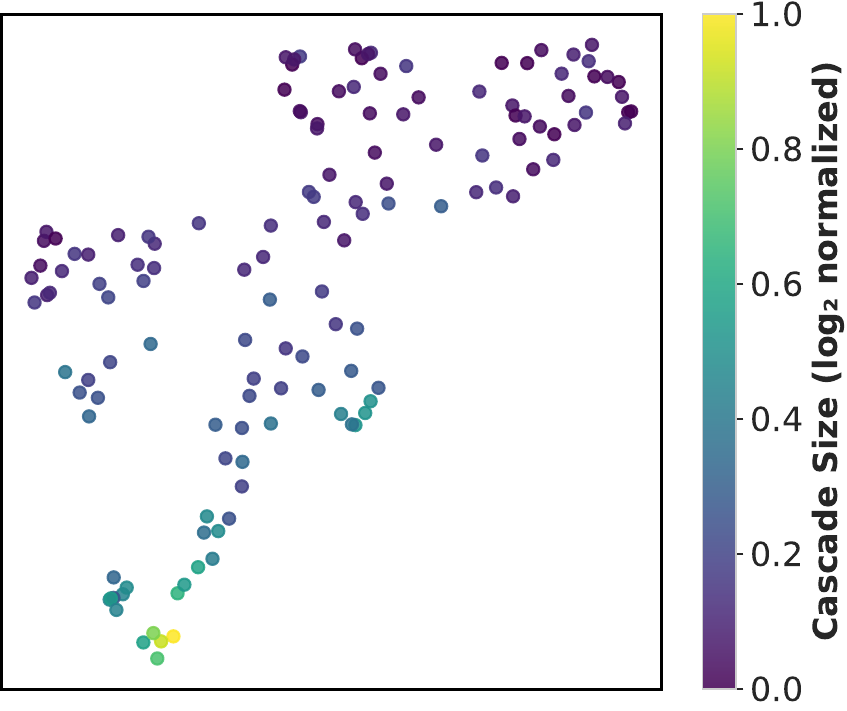}
        \caption*{\scriptsize (b) T3MAL w TTT.}
        \label{macro_umap_after}
    \end{subfigure}
    \caption{UMAP visualization of macroscopic cascade representations learned by T3MAL on the Android dataset, without and with test-time training. Each point corresponds to one of the 136 test cascades. Darker colors indicate larger final cascade sizes (log-scaled).}
    \label{fig:visualization_comparison}
    \vspace{-3mm}  
\end{figure}

\begin{figure*}[t]
    \centering

    \begin{subfigure}[b]{0.48\textwidth}
        \includegraphics[width=\linewidth]{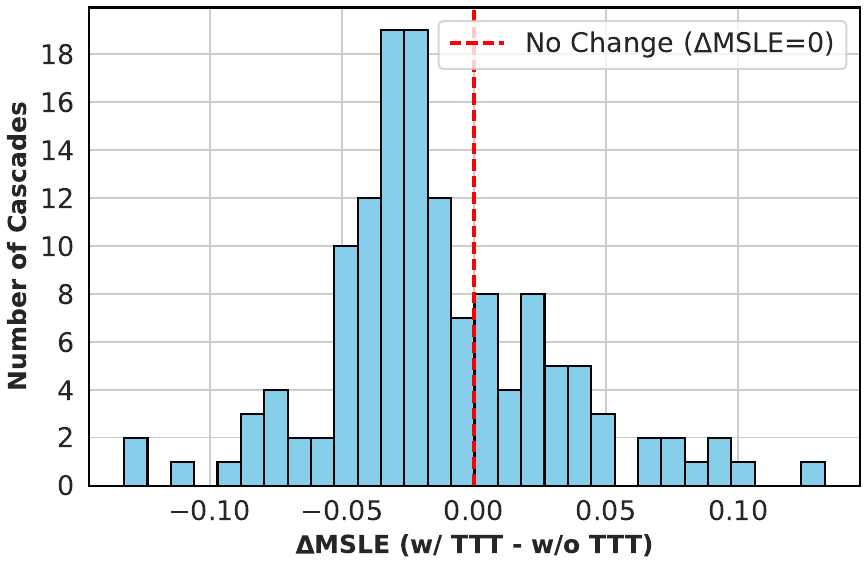}
        \caption{\small Jointly trained model with TTT (vanilla TTT).}
        \label{fig:vanilla-ttt}
    \end{subfigure}
    \hfill
    \begin{subfigure}[b]{0.48\textwidth}
        \includegraphics[width=\linewidth]{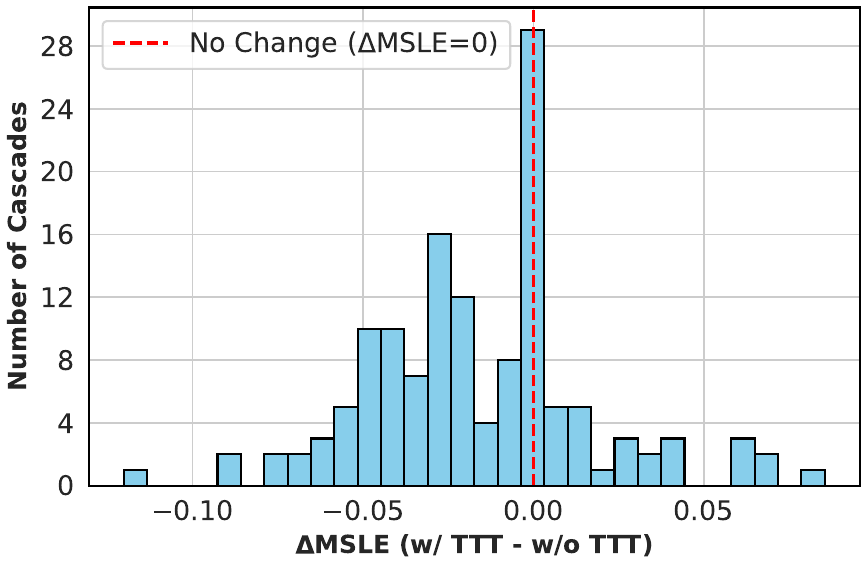}
        \caption{\small Meta-trained model with TTT (T3MAL).}
         \label{fig:t3mal-ttt}
    \end{subfigure}

    \caption{
        Histograms of $\Delta$MSLE per test cascade on the Android dataset, where $\Delta$MSLE is defined as MSLE\textsubscript{w/ TTT} $-$ MSLE\textsubscript{w/o TTT}. 
        A positive value indicates performance degradation after TTT, i.e., catastrophic forgetting. 
        Subfigures (a) and (b) correspond to vanilla TTT and our T3MAL method, respectively.
    }
    \label{fig:delta-msle-hists}
    \vspace{-2mm}  
\end{figure*}

\vspace{-0.4em}
\subsection{Visualization} \label{subsec:visualization}

To illustrate how test-time training improves model generalization, we visualize the macroscopic cascade representations learned by the feature encoder on the Android dataset, with and without TTT, as shown in Fig.~\ref{fig:visualization_comparison}. With TTT (right), the representations are more discriminative and exhibit stronger semantic structure: cascades with similar final sizes form tighter clusters, whereas those without TTT (left) appear more scattered. This improvement stems from the self-supervised learning during the meta-auxiliary testing phase, which adapts the feature encoder to better capture diffusion patterns specific to unseen test cascades. The visualization aligns with quantitative improvements in macroscopic prediction (MSLE), providing intuitive evidence that TTT enhances generalization by refining the representation space to better align with cascade dynamics at test time.

\vspace{-0.5em}
\subsection{Empirical Evidence of Catastrophic Forgetting} \label{subsec:analysis of catastrophic forgetting}

To empirically demonstrate that naively applying TTT may lead to undesirable catastrophic forgetting, we compare vanilla TTT and our proposed T3MAL on the Android dataset for the macroscopic prediction task. As shown in Fig.~\ref{fig:delta-msle-hists}, we plot $\Delta$MSLE for each test cascade, where $\Delta$MSLE = MSLE\textsubscript{w/ TTT} $-$ MSLE\textsubscript{w/o TTT}. We observe that a portion of test cascades exhibit $\Delta$MSLE $>$ 0, indicating that TTT sometimes causes catastrophic forgetting and degrades prediction performance.

Specifically, in Fig.~\ref{fig:vanilla-ttt}, although vanilla TTT reduces the average MSLE (from 0.144 to 0.130), it results in performance degradation on 31.62\% of test cascades. This is because the jointly trained model lacks foresight for future adaptation objective, requiring multiple gradient updates (four in this case) during TTT, which increases the risk of overfitting to the auxiliary task and forgetting previously learned knowledge. A similar issue is observed on the Twitter dataset: as shown in Table \ref{table: macroscopic_ablation_study_results} (Row 3 vs. Row 5), naively applying TTT even leads to overall performance degradation on the test set.

In contrast, Fig.~\ref{fig:t3mal-ttt} shows that T3MAL achieves a larger MSLE reduction (from 0.131 to 0.116) with two gradient steps, while also lowering the degradation ratio to 22.06\%. This demonstrates that T3MAL offers a better initialization for TTT and effectively mitigates catastrophic forgetting, thanks to its meta-auxiliary training and adaptor design. 

\vspace{-0.4em}
\subsection{Complexity and Efficiency Analysis}
\label{subsec:complexity and efficiency analysis}

\begin{figure}[t]
\centering
\includegraphics[width=\columnwidth]{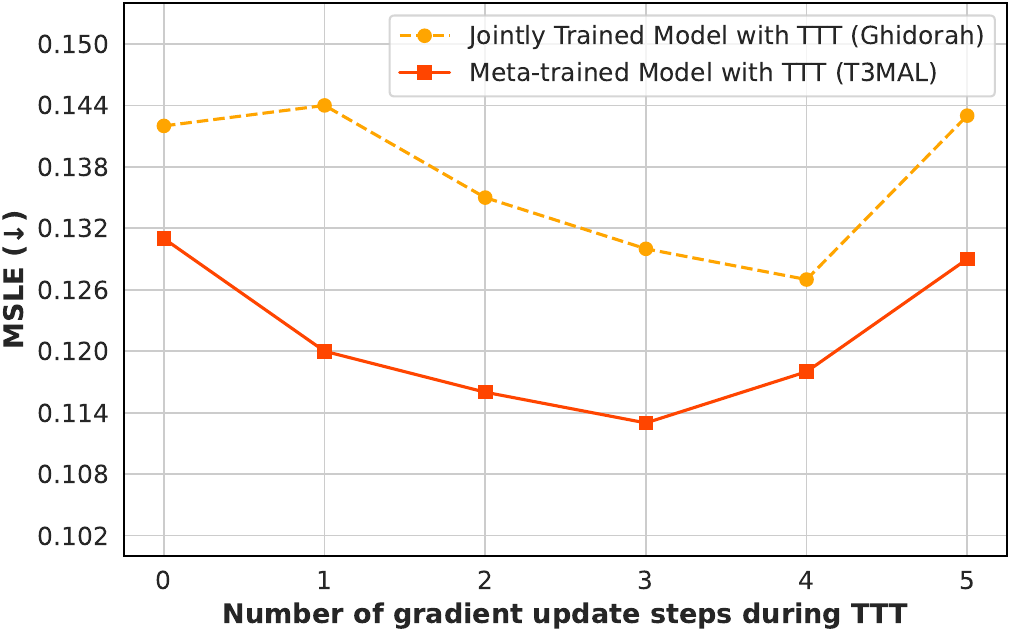} 
\caption{Performance of Ghidorah and T3MAL w.r.t update steps $\delta$ on the Android dataset, where $\delta=0$ is the initial performance of the jointly trained and meta-trained models without TTT.}
\vspace{-3mm}
\label{fig:Gradient update steps}
\vspace{-2mm}  
\end{figure}

For completeness, \textit{\textbf{Appendix H}} outlines the theoretical complexity of each T3MAL component, while this section reports empirical training and inference costs.

As shown in Fig. \ref{fig:Gradient update steps}, directly applying TTT to a jointly trained model, as in Ghidorah\cite{zhu2025ghidorah}, often results in suboptimal performance. This is because the model lacks foresight for TTT and cannot anticipate the number of update steps required, which vary across test cascades and are often substantial. Insufficient updates fail to fully adapt to the target distribution, while excessive updates risk overfitting and catastrophic forgetting. T3MAL addresses this by simulating TTT during training through meta-auxiliary learning, enabling the model to learn how to adapt effectively within a fixed number of updates and providing a better initialization.
Results in Fig.~\ref{fig:Gradient update steps} show that T3MAL outperforms Ghidorah with just one or two updates, achieving faster and more accurate adaptation. 

On the Android dataset, T3MAL takes $\sim$12.5 seconds per epoch for joint training (batch size 64) and $\sim$721.03 seconds per epoch for meta-auxiliary training ($\mathcal{T} = 5$, $\delta = 2$). At inference time, executing TTT with the same $\delta$ takes 16.4 seconds over the entire test set. In comparison, Ghidorah requires 9.98 seconds per epoch for joint training and 37.9 seconds for TTT with $\delta = 4$. While T3MAL incurs additional training overhead due to meta-auxiliary learning, this is a one-time offline cost. In return, it enables much faster and more accurate test-time adaptation with just 1–2 gradient steps. This reduction in $\delta$ greatly reduces the computational cost during inference, which is especially beneficial for large-scale test sets, since the cost increases linearly with $\delta$. 

As a result, the overall cost increase remains modest when considering both training and inference. Given the substantial improvements in predictive performance, this trade-off is well justified.

\section{RELATED WORK}
Our work relates to several research fields, including IDP, TTT, and meta-learning. Due to space limitations, we provide a brief overview here, with more detailed discussion available in \textit{\textbf{Appendix D}}.

\textit{Information diffusion prediction} aims to explore how information spreads over time based on observed user interactions. It is generally categorized into macroscopic prediction, which estimates a cascade’s future popularity, and microscopic prediction, which forecasts individual user engagements \cite{li2023grass,wang2023multiscale}. 
While many existing works target one of these tasks in isolation \cite{xu2021casflow,sun2022ms,wang2022cascade,feng2022h, cheng2023enhancing, lu2023continuous,ji2023community}, recent efforts have explored unified frameworks to address both tasks jointly \cite{chen2019information,yang2021full,jiao2024enhancing}. Despite their promising results, these methods often overlook distribution shifts between training and test cascades, which hinders their generalization and results in
unsatisfactory prediction performance on test data.

The \textit{test-time training} paradigm aims to adapt a trained model to each test instance before prediction using a self-supervised auxiliary task. TTT has gained increasing attention in various domains, including computer vision \cite{liu2021ttt++,hatem2023point,wang2024heterogeneous}, graph learning \cite{liu2023flood,zhang2024fully}, and rumor detection \cite{zhang2024t3rd}, where it has shown strong potential for handling distribution shifts. However, its effectiveness largely depends on the design of auxiliary task \cite{gandelsman2022test}, and if not properly constrained, it may negatively impact the primary task performance. Therefore, designing a robust and effective TTT framework for IDP tasks is non-trivial.

Meta-learning, often known as “learning to learn,” aims to extract transferable knowledge from related tasks to enable fast adaptation to new tasks with few examples. Existing methods can be broadly categorized into metric-based \cite{sung2018learning}, model-based \cite{santoro2016meta}, and optimization-based \cite{finn2017model, ravi2017optimization, sun2019meta} methods. Among them, Model-Agnostic Meta-Learning (MAML)~\cite{finn2017model} is widely adopted for its ability to learn a generalizable initialization through nested optimization \cite{park2020fast,soh2020meta}. 
Recent advances have extended meta-learning to auxiliary learning. MAXL~\cite{liu2019self} introduces a meta-auxiliary learning framework that jointly trains primary and auxiliary tasks, where a label-generation network is optimized to produce auxiliary labels that benefit primary task performance. 
Our approach draws inspiration from MAXL but differs in two key aspects. First, instead of generating auxiliary labels, we adopt BYOL as a self-supervised auxiliary task. Second, we employ MAML-style bi-level optimization to simulate the TTT process and explicitly couple the auxiliary objective with primary task performance, allowing the model to learn how to balance them for effective TTT.

\vspace{-0.4em}
\section{Conclusion}
In this paper, we proposed T3MAL, the first TTT-based learning framework for robust multi-scale information diffusion prediction under distribution shifts. It is designed to flexibly adapt a trained model to each test instance during inference, thereby improving generalization to unseen test data. Compared to vanilla TTT methods, T3MAL enables fast and accurate test-time adaptation through two innovative mechanisms: a novel meta-auxiliary learning scheme to learn better weight initialization for TTT and a lightweight adaptor to mitigate catastrophic forgetting. Extensive experiments demonstrate that our framework consistently outperforms state-of-the-art methods. 



\vspace{-0.4em}
\section*{Acknowledgments}
This work is supported by the National Key Research and
Development Program (Grant No. 2023YFC3303800).

\section*{Appendix}

\section*{A. Notation Summary}

\begin{table}[ht]
\centering
\caption{NOTATIONS USED IN THE PAPER}
\renewcommand{\arraystretch}{1.5} 
\begin{tabular}{@{}p{2cm} !{\vrule width 0.6pt} p{5.5cm}@{}} 
\toprule
\textbf{\small Notation} & \textbf{\small Description} \\ \midrule
{\fontfamily{ppl}\selectfont \textit{\small $\mathcal{C} = \{c_i\}_{i=1}^{M}$}} & {\fontfamily{ppl}\selectfont \small A set of diffusion cascades. Each $c_i$ is a cascade sequence.} \\
{\fontfamily{ppl}\selectfont \textit{\small $\mathcal{U} = \{u_i\}_{i=1}^{N}$}} & {\fontfamily{ppl}\selectfont \small A set of all users.} \\
{\fontfamily{ppl}\selectfont \textit{\small $\mathcal{G}_S = (\mathcal{U}, \mathcal{E})$}} & {\fontfamily{ppl}\selectfont \small Social graph with users $\mathcal{U}$ and edges $\mathcal{E}$.} \\
{\fontfamily{ppl}\selectfont \textit{\small $\mathcal{G}^t_D = (\mathcal{U}^t, \mathcal{E}^t)$}} & {\fontfamily{ppl}\selectfont \small Diffusion hypergraph for time interval $t$, where $\mathcal{U}^t \subseteq \mathcal{U}$.} \\
{\fontfamily{ppl}\selectfont \textit{\small $\mathbf{X}_S \in \mathbb{R}^{N \times d}$}} & {\fontfamily{ppl}\selectfont \small User embeddings from social graph.} \\
{\fontfamily{ppl}\selectfont \textit{\small $\mathbf{X}_D \in \mathbb{R}^{N \times d}$}} & {\fontfamily{ppl}\selectfont \small User embeddings from diffusion hypergraphs.} \\
{\fontfamily{ppl}\selectfont \textit{\small $a_i,\tilde{a}_i$}} & {\fontfamily{ppl}\selectfont \small Two augmented views of cascade $c_i$.}\\
{\fontfamily{ppl}\selectfont \textit{\small $\mathcal{F}$}} & {\fontfamily{ppl}\selectfont \small Generic feature encoder.} \\
{\fontfamily{ppl}\selectfont \textit{\small $\mathcal{A}_\theta$}} & {\fontfamily{ppl}\selectfont \small Adaptor for customizing encoder $\mathcal{F}$.}\\
{\fontfamily{ppl}\selectfont \textit{\small $\mathcal{Q}_\theta,\mathcal{P}_\theta$}} & {\fontfamily{ppl}\selectfont \small Projector and predictor in the BYOL auxiliary network.} \\
{\fontfamily{ppl}\selectfont \textit{\small $\psi_p,\psi_m$}} & {\fontfamily{ppl}\selectfont \small Macro/micro prediction heads.} \\
{\fontfamily{ppl}\selectfont \textit{\small $\alpha,\beta$}} & {\fontfamily{ppl}\selectfont \small Inner/meta learning rates.} \\
{\fontfamily{ppl}\selectfont \textit{\small $\delta$}} & {\fontfamily{ppl}\selectfont \small Gradient steps for test-time adaptation.}  \\
{\fontfamily{ppl}\selectfont \textit{\small $\lambda, \gamma$}} & {\fontfamily{ppl}\selectfont \small Weights for loss terms.}\\
{\fontfamily{ppl}\selectfont \textit{\small $\theta, \xi$}} & {\fontfamily{ppl}\selectfont \small Params of online and target networks.}\\
{\fontfamily{ppl}\selectfont \textit{\small $\mathcal{T}$}} & {\fontfamily{ppl}\selectfont \small Meta-training batch size.}   
\\ \bottomrule
\end{tabular}
\label{table:notations}
\end{table}

\section*{B. Dataset Details}

We conduct experiments on three publicly available real-world datasets: Android, Twitter, and Douban.

The \textbf{Android} dataset \cite{sankar2020inf}, collected from StackExchange, encompasses discussions centered on Android-related topics. The social network is constructed based on user interactions such as questions, answers, comments, and upvotes. Information cascades are represented as chronologically ordered sequences of posts associated with the same tag.

The \textbf{Twitter} dataset \cite{hodas2014simple} contains tweets with embedded URLs collected during October 2010. Each URL is treated as an information item that propagates among users. The social network is constructed based on the follower–followee relationships among users on Twitter.

The \textbf{Douban} dataset \cite{zhong2012comsoc} is collected from a Chinese social platform where users share their reading activities and follow updates from others. Each book is treated as an information item, and a user is considered “infected” upon reading it. The social network is constructed from online contact relations and offline co-occurrence in social gatherings.


Detailed dataset statistics are reported in the main text and omitted here for brevity.

\section*{C. Baseline Details}

We benchmark our models against sixteen representative baseline models.

\textbf{Macroscopic prediction models:}
\begin{itemize}
\item \textbf{DeepCas} \cite{li2017deepcas} represents a cascade graph as a set of paths sampled via random walks, and encodes them using a BiGRU with an attention mechanism to learn the representation of the entire cascade graph for prediction.
\item \textbf{DeepHawkes} \cite{cao2017deephawkes} transforms each cascade into a set of retweet paths and leverages a GRU with sum pooling to encode them, aiming to model the interpretable factors of the Hawkes process.
\item \textbf{CasCN}~\cite{chen2019information} leverages both structural and temporal information by sampling the cascade graph into a sequence of sub-cascade graphs. It employs a multi-directional GCN to learn local structures within each subgraph, and an LSTM to capture the evolution of the cascade structure for prediction~\footnote{https://github.com/ChenNed/CasCN}.
\item \textbf{CasFlow}~\cite{xu2021casflow} models hierarchical cascade diffusion by learning local/global structure via graph wavelets and matrix factorization, and captures uncertainty at both node and cascade levels using variational autoencoders and normalizing flows.
\end{itemize}

\textbf{Microscopic prediction models:}
\begin{itemize}
\item \textbf{TopoLSTM} \cite{wang2017topological} models diffusion as a dynamic DAG and extends the standard LSTM to learn sender embeddings for active nodes, enabling accurate activation prediction.
\item \textbf{NDM}~\cite{yang2019neural} targets general microscopic cascade modeling without requiring an explicit diffusion graph. It uses multi-head attention to capture user influence and CNN to aggregate active user representations for next-user prediction~\footnote{https://github.com/albertyang33/NeuralDiffusionModel}. 
\item \textbf{Inf-VAE}~\cite{sankar2020inf} is a VAE framework that integrates structure-preserving social embeddings and position-encoded temporal embeddings via a co-attentive fusion network for prediction~\footnote{https://github.com/aravindsankar28/Inf-VAE}.
\item \textbf{DyHGCN}~\cite{yuan2021dyhgcn} constructs a dynamic heterogeneous graph to jointly model social and diffusion relations, incorporates temporal signals, and learns users' dynamic preferences for diffusion prediction.
\item \textbf{MS-HGAT}~\cite{sun2022ms} models static user dependencies with GCN and dynamic interactions with sequential hypergraph attention, and uses a memory module to capture evolving user preferences for prediction~\footnote{https://github.com/slingling/MS-HGAT}.
\item \textbf{CE-GCN}~\cite{wang2022cascade} builds a heterogeneous graph with user and cascade nodes linked by social, diffusion, and enhancement edges, and employs message passing and cascade-specific aggregation to refine user representations for diffusion prediction.
\item \textbf{RotDiff} \cite{qiao2023rotdiff} RotDiff leverages Lorentzian embeddings and hyperbolic rotation transformations to model asymmetric, hierarchical social factors for improved information diffusion prediction.
\end{itemize}

\textbf{Multi-scale prediction models:}
\begin{itemize}
\item \textbf{FOREST}~\cite{yang2021full} is a full-scale diffusion prediction framework that combines a GRU-based microscopic model with structural context from the social graph. It further adopts cascade simulation with reinforcement learning to align microscopic and macroscopic objectives.
\item \textbf{MINDS}~\cite{jiao2024enhancing} captures user preferences through sequential hypergraphs and social homophily, and learns shared representations across microscopic and macroscopic tasks via a shared LSTM with adversarial and orthogonality constraints.
\end{itemize}

\textbf{DG and UDA models:}
\begin{itemize}
\item \textbf{DANN}~\cite{ganin2016domain} performs unsupervised domain adaptation by learning label-discriminative but domain-invariant features, using a gradient reversal layer to adversarially train the feature extractor against a domain classifier.
\item \textbf{MLDG} \cite{li2018learning} is a meta-learning method for domain generalization that simulates domain shift by splitting source domains into meta-train and meta-test subsets within each mini-batch.
\item \textbf{SFA}~\cite{li2021simple} improves domain generalization by injecting stochastic noise into feature embeddings, using either data-independent Gaussian noise or class-conditional adaptive noise to simulate domain shifts without requiring domain labels.
\item \textbf{CLUDA}~\cite{ozyurt2023contrastive} is a contrastive learning framework for unsupervised domain adaptation of time series. It learns domain-invariant representations via semantic-preserving augmentations, adversarial training, and nearest-neighbor contrastive learning.
\end{itemize}

\section*{D. RELATED WORK}
Our work is relevant to several research fields, including information diffusion prediction, test-time training and meta-auxiliary learning.

\subsection{Information Diffusion Prediction}
Research on information diffusion prediction (IDP) aims to uncover underlying diffusion patterns by using observed cascade data and structural signals (e.g., social networks and cascade graphs). The goal is to model complex diffusion dependencies among users and predict how a cascade will evolve over time \cite{gao2019popularity}. Based on the prediction granularity, existing IDP models are generally categorized into two types: \textit{macroscopic prediction} and \textit{microscopic prediction}. The former, also known as \textit{popularity prediction}, estimates the overall reach of a cascade, such as the total number of retweets. The latter focuses on individual behavior, predicting whether specific users will participate in the diffusion process and estimating their likelihood of doing so.


For macroscopic prediction, related work falls into three main categories. Early studies focus on feature-based methods that engineer handcrafted temporal, structural, and content features to represent cascades \cite{cheng2014can, shulman2016predictability, chen2020event}, but these approaches heavily depend on feature quality and often lack generalizability. Generative methods model diffusion with Hawkes point processes \cite{mishra2016feature}, offering interpretability but with limited predictive power. 
More recently, deep learning-based approaches decompose cascade graphs into diffusion paths, encode them using sequential models such as RNNs or LSTMs, and aggregate the representations for prediction \cite{li2017deepcas, cao2017deephawkes}.
While these methods represent progress, they make limited use of the structural information inherent in cascade graphs. To address this, recent efforts focus on more effective modeling of cascade graphs \cite{chen2019information, xu2021casflow, tang2021fully, lu2023continuous, cheng2024information}. For example, CasFlow \cite{xu2021casflow} jointly models social and cascade graphs to capture richer structural and temporal patterns, while CTCP \cite{lu2023continuous} further considers the dynamic evolution of cascades and their interactions. These approaches achieve significant improvements in predictive performance.

Microscopic prediction closely resembles macroscopic prediction, differing primarily in the output layer. 
Currently, deep learning-based approaches dominate this task, offering the most advanced solutions. Early efforts primarily treat diffusion cascades as sequences, leveraging attention mechanisms to capture diffusion dependencies among users \cite{yang2019neural,wang2019hierarchical}. More recently, researchers propose to first encode the available graph structure through graph neural networks to obtain user representations, and then model the diffusion cascades based on temporal information\cite{sankar2020inf, yuan2021dyhgcn, wang2022cascade, sun2022ms}. For example, Inf-VAE \cite{sankar2020inf} jointly models the social network and temporal influence via a variational autoencoder and co-attention mechanism. DyHGCN \cite{yuan2021dyhgcn} and CE-GCN \cite{wang2022cascade} construct heterogeneous graphs that integrate diffusion and social edges to learn more informative user representations. Additionally, MS-HGAT \cite{sun2022ms} adopts a hypergraph structure to better capture global user interactions.

Recent advances have introduced multi-scale information diffusion models to jointly address microscopic and macroscopic prediction tasks \cite{yang2021full, chen2019information, jiao2024enhancing}. For example, FOREST \cite{yang2021full} performs macroscopic prediction using a modified microscopic model, in which cascade simulation is integrated into a reinforcement learning framework. MINDS \cite{jiao2024enhancing}, inspired by multi-task learning, employs distinct modules to extract shared and task-specific features, while leveraging adversarial training and orthogonality constraints to mitigate feature interference. Despite their promise, these methods largely overlook the challenge of out-of-distribution generalization.

\subsection{Test-Time Training}
The test-time training paradigm aims to adapt a trained model to each test instance before prediction using a self-supervised auxiliary task. TTT has gained increasing attention in various domains, including computer vision \cite{liu2021ttt++,hatem2023point,wang2024heterogeneous}, graph learning \cite{liu2023flood,zhang2024fully}, and rumor detection \cite{zhang2024t3rd}, where it has shown strong potential for handling distribution shifts. However, its effectiveness largely depends on the design of auxiliary task \cite{gandelsman2022test}, and if not properly constrained, it may negatively impact the primary task performance. Therefore, designing a robust and effective TTT framework for IDP tasks is non-trivial.
\subsection{Meta Learning}

Meta-learning, often known as “learning to learn,” aims to extract transferable knowledge from related tasks to enable fast adaptation to new tasks with few examples. Existing methods can be broadly categorized into metric-based \cite{koch2015siamese,sung2018learning}, model-based \cite{santoro2016meta}, and optimization-based \cite{finn2017model, ravi2017optimization, sun2019meta} methods. Among them, Model-Agnostic Meta-Learning (MAML)~\cite{finn2017model} is widely adopted for its ability to learn a generalizable initialization through nested optimization, where task-specific adaptation is performed in the inner loop and meta-level updates are applied in the outer loop \cite{park2020fast,soh2020meta}.

Recent advances have extended meta-learning to auxiliary learning. MAXL~\cite{liu2019self} introduces a meta-auxiliary learning framework that jointly trains primary and auxiliary tasks, where a label-generation network is optimized to produce auxiliary labels that benefit primary task performance. This allows the model to discover useful auxiliary labels without requiring manually defined subcategories. Our approach draws inspiration from MAXL but differs in two key aspects. First, instead of learning auxiliary labels, we adopt BYOL as a self-supervised auxiliary task. Second, we employ MAML-style bi-level optimization to learn the optimal trade-off between the primary and auxiliary tasks. By jointly setting them as meta-objectives, our method promotes synergy between them and enables effective test-time training.

\section*{E. Algorithm of Meta-Auxiliary Training}
In this part, we provide the pseudocode for the meta-auxiliary training procedure of the proposed T3MAL model in Algorithm~\ref{alg:meta_training}. 

\begin{algorithm}[t]
\caption{Meta-Auxiliary Training}
\label{alg:meta_training}
\REQUIRE Training data $\mathcal{D}^{tr}$, gradient updates $\delta$, batch size $\mathcal{T}$, inner learning rate $\alpha$, meta-learning rate $\beta$\\
\textbf{Output}: meta-auxiliary learned parameters $\theta^*$

\begin{algorithmic}[1]
\State Initialize the model with jointly trained weights $\theta$
\While{\textit{not converged}}
    \State Sample a batch of training cascades $\{c_i\}_{i=1}^\mathcal{T}$
    \For{\textit{each cascade} $c_i$}
        \State Generate augmentations $a_i$ and $\tilde{a}_i$ 
        \For{\textit{inner step} $= 1, 2, \dots, \delta$}
            \State Calculate $r_{\phi_i}^i$, $\tilde{r}_{\phi_i}^i$, $z_{\theta}^{i}$, and $\tilde{z}_{\theta}^{i}$
            \State Update task-specific model $\phi_i$:
            \State $\phi_i \leftarrow \theta - \alpha \nabla_{\theta} \mathcal{L}_{Aux}(c_i;\theta)$ \Comment{Eq.\ 14}
        \EndFor
    \EndFor
    \State Update meta-model $\theta$:
    \State $\theta \leftarrow \theta - \beta \nabla_\theta \sum_{i=1}^{\mathcal{T}} \mathcal{L}_{Meta}(c_i;\phi_i)$ \Comment{Eq.\ 16}
\EndWhile
\State \textbf{Return} $\theta^*$
\end{algorithmic}
\end{algorithm}

\section*{F. Ablation Study on Microscopic Task}

For completeness, we also conducted an extensive ablation study on the microscopic prediction task to validate the contribution of each module within T3MAL. Results are summarized in Table~\ref{table: microscopic_ablation_study_results}.

\begin{table}[t]
    \caption{Ablation study results on Android and Twitter datasets for microscopic prediction task, evaluated by the Hits@100 metric.}
    \label{table: microscopic_ablation_study_results}
    \begin{center}
    \begin{small}
    \resizebox{1.0\columnwidth}{!}{
    \begin{tabular}{p{2cm}lcc}
    \toprule
    \textbf{Type} & \textbf{Variant} & \textbf{Android} & \textbf{Twitter} \\
    \midrule
    \multirow{1}{*}{\centering Primary Only} & - Micro & 0.2747 & 0.5793 \\
    \midrule
    \multirow{2}{*}{\centering Joint Training} & - Micro + BYOL & 0.2828 & 0.5848 \\
    & - Micro + Macro + BYOL & 0.3040 & 0.6013 \\
    \midrule
    \multirow{2}{*}{\centering Meta-Training} & - Micro + BYOL + MT & 0.3021 & 0.6026 \\
    & - Micro + Macro + BYOL + MT & 0.3237 & 0.6177 \\
    \midrule
    \multirow{3}{*}{\centering Meta-Testing} & - Micro + BYOL + TTT & 0.2914 & 0.5812 \\
    & - Micro + Macro + BYOL + TTT & 0.3133 & 0.5972 \\
    & - Micro + BYOL + MT + TTT & 0.3255 & 0.6224 \\
    \midrule
    \multirow{2}{*}{\centering Auxiliary Task} & - Contrastive Learning & 0.3189 & 0.6164 \\
    & - MAE-based Self-Reconstruction & 0.3252 & 0.6216 \\
    \midrule
    \multirow{1}{*}{\centering BYOL Variant} & - w/o Adaptor & 0.3311 & 0.6270 \\
    \midrule
    \addlinespace
    \multirow{1}{*}{\centering \textbf{T3MAL}} & \textbf{All} & \textbf{0.3372} & \textbf{0.6335} \\
    \bottomrule
    \end{tabular}
    }
    \end{small}
    \end{center}
\end{table}


\section*{G. Hyperparameter Sensitivity Analysis}

\begin{figure}[t]
\centering
\includegraphics[width=0.9\columnwidth]{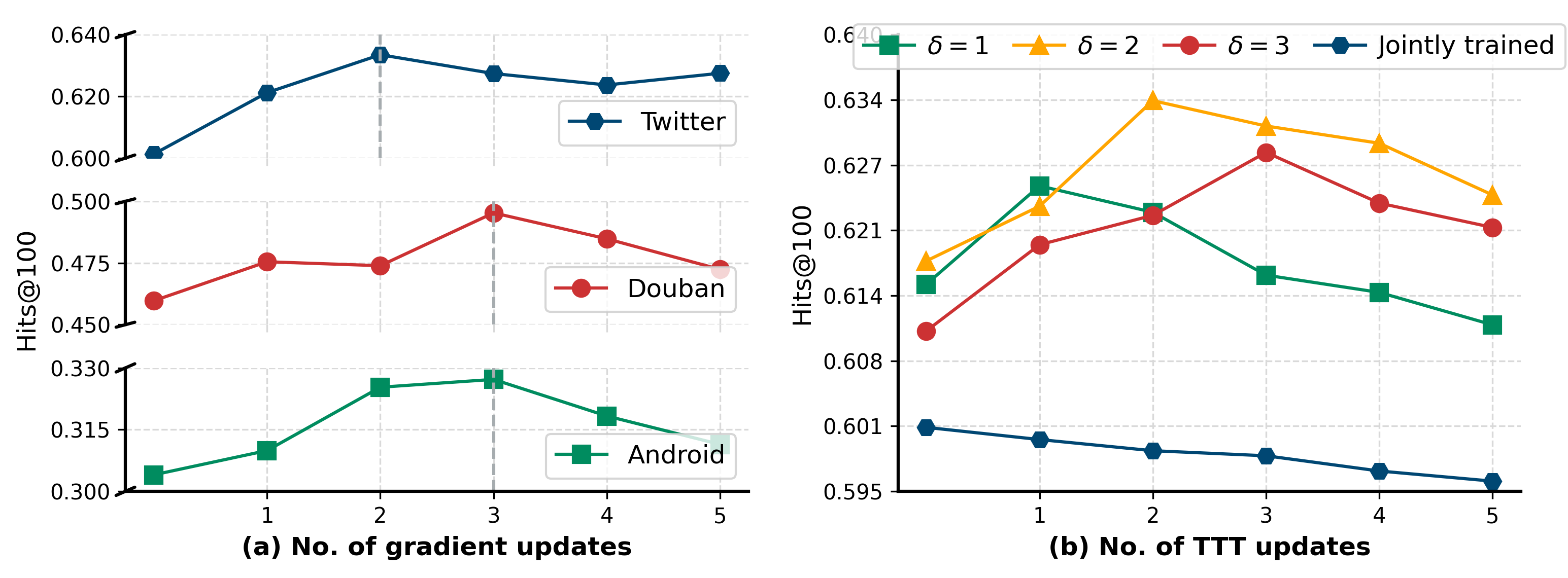} 
\caption{Gradient update steps Analysis.}
\vspace{-3mm}
\label{fig：Gradient update steps}
\end{figure}

\textit{\textbf{1) Sensitivity Analysis for Gradient Update Steps $\bm{\delta}$.}}
The effect of inner-loop update steps $\delta$ (as defined in Algorithm~\ref{alg:meta_training}) is evaluated across three benchmark datasets. When $\delta=0$, we evaluate the jointly trained models optimized via Eq.(11). As shown in Fig.~\ref{fig：Gradient update steps}(a), the performance of T3MAL improves with more gradient updates and peaks around $\delta=2$ or $\delta=3$, depending on the dataset. Beyond this point, additional updates lead to diminishing returns or even performance drops. This phenomenon likely stems from inner-loop overfitting to task-irrelevant cascade details while forgetting jointly trained knowledge. Notably, we adopt the same $\delta$ for both meta-auxiliary training and test-time training. To validate this design choice, we conduct experiments on the Twitter dataset using meta-models trained with $\delta = \{1, 2,3\}$ and the jointly trained model, while varying $\delta$ during TTT. Fig.~\ref{fig：Gradient update steps}(b) shows that performance is optimal when TTT employs the same $\delta$ used during meta-auxiliary training. This observation is intuitively reasonable: meta-auxiliary training essentially simulates the TTT process, and the meta-trained models with a specific $\delta$ value have learned how to perform effective test-time adaptation with $\delta$-step updates. Moreover, directly applying TTT to a jointly trained model—without meta-auxiliary training—degrades performance rather than improving it, underscoring the the necessity of meta-auxiliary training.

\begin{figure}[t]
\centering
\includegraphics[width=0.9\columnwidth]{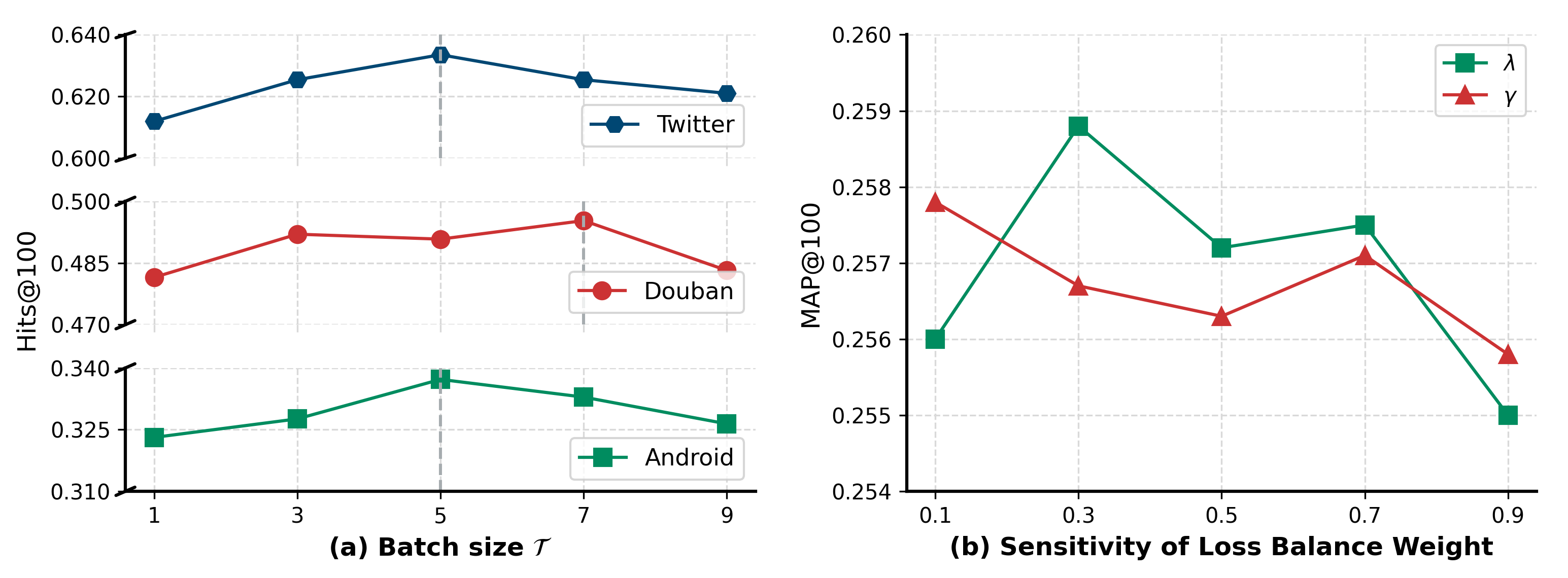} 
\caption{Batch size and Balance Weight.}
\vspace{-3mm}
\label{fig: Batch size & Balance Weight}
\end{figure}

\textit{\textbf{2) Sensitivity Analysis for Batch Size $\bm{\mathcal{T}}$}.} 
As shown in Fig.~\ref{fig: Batch size & Balance Weight}(a), we examine the effect of batch size $\mathcal{T} = \{1, 3, 5, 7, 9\}$ during meta-auxiliary training. Overall, model performance tends to improve with increasing $\mathcal{T}$, reaching its peak at $\mathcal{T} = 5$ on Twitter and Android, and at $\mathcal{T} = 7$ on Douban. These results suggest that moderately larger batch sizes can help mitigate overfitting to individual cascades. However, further increasing $\mathcal{T}$ beyond the peak yields diminishing or even adverse returns, indicating a saturation point.


\textit{\textbf{3) Sensitivity Analysis for Loss Balance Weight $\bm{\lambda}$ and $\bm{\gamma}$.}}
We evaluate the sensitivity of T3MAL to two hyperparameters on the Twitter dataset: the balancing weight $\lambda$ for the two primary tasks (Eq.(10)) and the loss weight $\gamma$ for the auxiliary task (Eq.(11), (16)). As illustrated in Fig.~\ref{fig: Batch size & Balance Weight}(b), T3MAL maintains robust performance when both hyperparameters fall within moderate ranges, demonstrating its resilience to hyperparameter variations.


\section*{H. Complexity Analysis}

To evaluate the efficiency of the proposed method, we analyze the time complexity of its main components.

\begin{itemize}
\item \textbf{Complexity for Learning User Representations:} 
We adopt a multi-layer GCN to model the social graph, with a time complexity of $\mathcal{O}(L_s(|\mathcal{E}|d + N d^2))$, where $L_s$ is the number of GCN layers. For the temporal diffusion hypergraphs with 
$T$ intervals, a single-layer HGNN is applied to each interval $t$, resulting in a total complexity of $\mathcal{O}(T(|\mathcal{E}^t| + |\mathcal{U}^t|)d^2)$. The overall cost scales linearly with the number of users and edges in both graphs.
\item \textbf{Complexity for Generic Feature Encoder and Adaptor:} 
The generic encoder consists of a shared LSTM and two task-specific LSTMs, with a time complexity of $\mathcal{O}(3 L_c d^2)$, where $L_c$ is the cascade length. The adaptor generates FiLM parameters via MLPs to adapt all LSTM weights and biases. Since the output size of the adaptor is proportional to the number of encoder parameters $\mathcal{O}(d^2)$, its computational complexity is $\mathcal{O}(d^3)$. 
\item \textbf{Complexity for Other Parts of T3MAL:} 
The projector and predictor in the auxiliary network, as well as the prediction heads in the primary network, are all implemented as MLPs. Their complexities depend on the hidden dimension $d$, while the microscopic prediction head additionally scales with the number of users $N$.
\end{itemize}


\bibliography{references}

\begin{thebibliography}{10}
\providecommand{\url}[1]{#1}
\csname url@samestyle\endcsname
\providecommand{\newblock}{\relax}
\providecommand{\bibinfo}[2]{#2}
\providecommand{\BIBentrySTDinterwordspacing}{\spaceskip=0pt\relax}
\providecommand{\BIBentryALTinterwordstretchfactor}{4}
\providecommand{\BIBentryALTinterwordspacing}{\spaceskip=\fontdimen2\font plus
\BIBentryALTinterwordstretchfactor\fontdimen3\font minus \fontdimen4\font\relax}
\providecommand{\BIBforeignlanguage}[2]{{%
\expandafter\ifx\csname l@#1\endcsname\relax
\typeout{** WARNING: IEEEtran.bst: No hyphenation pattern has been}%
\typeout{** loaded for the language `#1'. Using the pattern for}%
\typeout{** the default language instead.}%
\else
\language=\csname l@#1\endcsname
\fi
#2}}
\providecommand{\BIBdecl}{\relax}
\BIBdecl

\bibitem{xu2021casflow}
X.~Xu, F.~Zhou, K.~Zhang, S.~Liu, and G.~Trajcevski, ``Casflow: Exploring hierarchical structures and propagation uncertainty for cascade prediction,'' \emph{TKDE}, vol.~35, no.~4, pp. 3484--3499, 2021.

\bibitem{sun2022ms}
L.~Sun, Y.~Rao, X.~Zhang, Y.~Lan, and S.~Yu, ``Ms-hgat: memory-enhanced sequential hypergraph attention network for information diffusion prediction,'' in \emph{AAAI}, vol.~36, no.~4, 2022, pp. 4156--4164.

\bibitem{feng2022h}
S.~Feng, K.~Zhao, L.~Fang, K.~Feng, W.~Wei, X.~Li, and L.~Shao, ``H-diffu: hyperbolic representations for information diffusion prediction,'' \emph{TKDE}, vol.~35, no.~9, pp. 8784--8798, 2022.

\bibitem{bao2024popularity}
P.~Bao, R.~Yan, and C.~Yang, ``Popularity prediction via modeling temporal dependencies on dynamic evolution process,'' \emph{TKDE}, vol.~36, no.~11, pp. 6828--6838, 2024.

\bibitem{yang2021full}
C.~Yang, H.~Wang, J.~Tang, C.~Shi, M.~Sun, G.~Cui, and Z.~Liu, ``Full-scale information diffusion prediction with reinforced recurrent networks,'' \emph{TNNLS}, vol.~34, no.~5, pp. 2271--2283, 2021.

\bibitem{jiao2024enhancing}
P.~Jiao, H.~Chen, Q.~Bao, W.~Zhang, and H.~Wu, ``Enhancing multi-scale diffusion prediction via sequential hypergraphs and adversarial learning,'' in \emph{AAAI}, vol.~38, 2024, pp. 8571--8581.

\bibitem{quinonero2022dataset}
J.~Qui{\~n}onero-Candela, M.~Sugiyama, A.~Schwaighofer, and N.~D. Lawrence, \emph{Dataset shift in machine learning}.\hskip 1em plus 0.5em minus 0.4em\relax Mit Press, 2022.

\bibitem{jia2022hedan}
X.~Jia, J.~Shang, D.~Liu, H.~Zhang, and W.~Ni, ``Hedan: Heterogeneous diffusion attention network for popularity prediction of online content,'' \emph{KBS}, p. 109659, 2022.

\bibitem{yuan2021dyhgcn}
C.~Yuan, J.~Li, W.~Zhou, Y.~Lu, X.~Zhang, and S.~Hu, ``Dyhgcn: A dynamic heterogeneous graph convolutional network to learn users' dynamic preferences for information diffusion prediction,'' in \emph{ECML-PKDD}.\hskip 1em plus 0.5em minus 0.4em\relax Springer, 2020.

\bibitem{wang2023exploring}
J.~Wang and J.~Xie, ``Exploring the factors influencing users' learning and sharing behavior on social media platforms,'' \emph{Library Hi Tech}, vol.~41, no.~5, pp. 1436--1455, 2023.

\bibitem{ji2023community}
S.~Ji, X.~Lu, M.~Liu, L.~Sun, C.~Liu, B.~Du, and H.~Xiong, ``Community-based dynamic graph learning for popularity prediction,'' in \emph{SIGKDD}, 2023, pp. 930--940.

\bibitem{ziakas2024public}
P.~D. Ziakas and E.~Mylonakis, ``Public interest trends for covid-19 and pandemic trajectory: A time-series analysis of us state-level data,'' \emph{PLOS Digital Health}, vol.~3, p. e0000462, 2024.

\bibitem{fernando2013unsupervised}
B.~Fernando, A.~Habrard, M.~Sebban, and T.~Tuytelaars, ``Unsupervised visual domain adaptation using subspace alignment,'' in \emph{ICCV}, 2013, pp. 2960--2967.

\bibitem{long2018conditional}
M.~Long, Z.~Cao, J.~Wang, and M.~I. Jordan, ``Conditional adversarial domain adaptation,'' \emph{NeurIPS}, vol.~31, 2018.

\bibitem{ganin2016domain}
Y.~Ganin, E.~Ustinova, H.~Ajakan, P.~Germain, H.~Larochelle, F.~Laviolette, M.~March, and V.~Lempitsky, ``Domain-adversarial training of neural networks,'' \emph{JMLR}, vol.~17, pp. 1--35, 2016.

\bibitem{balaji2018metareg}
Y.~Balaji, S.~Sankaranarayanan, and R.~Chellappa, ``Metareg: Towards domain generalization using meta-regularization,'' \emph{NeurIPS}, vol.~31, 2018.

\bibitem{carlucci2019domain}
F.~M. Carlucci, A.~D'Innocente, S.~Bucci, B.~Caputo, and T.~Tommasi, ``Domain generalization by solving jigsaw puzzles,'' in \emph{CVPR}, 2019, pp. 2229--2238.

\bibitem{wang2022generalizing}
J.~Wang, C.~Lan, C.~Liu, Y.~Ouyang, T.~Qin, W.~Lu, Y.~Chen, W.~Zeng, and S.~Y. Philip, ``Generalizing to unseen domains: A survey on domain generalization,'' \emph{TKDE}, vol.~35, pp. 8052--8072, 2022.

\bibitem{sun2020test}
Y.~Sun, X.~Wang, Z.~Liu, J.~Miller, A.~Efros, and M.~Hardt, ``Test-time training with self-supervision for generalization under distribution shifts,'' in \emph{ICML}.\hskip 1em plus 0.5em minus 0.4em\relax PMLR, 2020, pp. 9229--9248.

\bibitem{chen2022ost}
L.~Chen, Y.~Zhang, Y.~Song, J.~Wang, and L.~Liu, ``Ost: Improving generalization of deepfake detection via one-shot test-time training,'' \emph{NeurIPS}, vol.~35, pp. 24\,597--24\,610, 2022.

\bibitem{Sain_2022_CVPR}
A.~Sain, A.~K. Bhunia, V.~Potlapalli, P.~N. Chowdhury, T.~Xiang, and Y.-Z. Song, ``Sketch3t: Test-time training for zero-shot sbir,'' in \emph{CVPR}, June 2022, pp. 7462--7471.

\bibitem{gandelsman2022test}
Y.~Gandelsman, Y.~Sun, X.~Chen, and A.~Efros, ``Test-time training with masked autoencoders,'' \emph{NeurIPS}, vol.~35, pp. 29\,374--29\,385, 2022.

\bibitem{zhu2025ghidorah}
W.~Zhu, C.~Li, L.~Zhang, S.~Wang, and X.~Zhang, ``Ghidorah: Towards robust multi-scale information diffusion prediction via test-time training,'' in \emph{AAAI}, vol.~39, 2025, pp. 13\,464--13\,472.

\bibitem{mccloskey1989catastrophic}
M.~McCloskey and N.~J. Cohen, ``Catastrophic interference in connectionist networks: The sequential learning problem,'' in \emph{Psychology of learning and motivation}.\hskip 1em plus 0.5em minus 0.4em\relax Elsevier, 1989, vol.~24, pp. 109--165.

\bibitem{grill2020bootstrap}
J.-B. Grill, F.~Strub, F.~Altch{\'e}, C.~Tallec, P.~Richemond, E.~Buchatskaya, C.~Doersch, B.~Avila~Pires, Z.~Guo, M.~Gheshlaghi~Azar \emph{et~al.}, ``Bootstrap your own latent-a new approach to self-supervised learning,'' \emph{NeurIPS}, pp. 21\,271--21\,284, 2020.

\bibitem{kipf2016semi}
T.~N. Kipf and M.~Welling, ``Semi-supervised classification with graph convolutional networks,'' in \emph{ICLR}, 2017.

\bibitem{cheng2023enhancing}
Z.~Cheng, W.~Ye, L.~Liu, W.~Tai, and F.~Zhou, ``Enhancing information diffusion prediction with self-supervised disentangled user and cascade representations,'' in \emph{CIKM}, 2023, pp. 3808--3812.

\bibitem{xie2022contrastive}
X.~Xie, F.~Sun, Z.~Liu, S.~Wu, J.~Gao, J.~Zhang, B.~Ding, and B.~Cui, ``Contrastive learning for sequential recommendation,'' in \emph{ICDE}.\hskip 1em plus 0.5em minus 0.4em\relax IEEE, 2022, pp. 1259--1273.

\bibitem{perez2018film}
E.~Perez, F.~Strub, H.~De~Vries, V.~Dumoulin, and A.~Courville, ``Film: Visual reasoning with a general conditioning layer,'' in \emph{AAAI}, vol.~32, no.~1, 2018.

\bibitem{lu2023continuous}
X.~Lu, S.~Ji, L.~Yu, L.~Sun, B.~Du, and T.~Zhu, ``Continuous-time graph learning for cascade popularity prediction,'' in \emph{IJCAI}, 2023, pp. 2224--2232.

\bibitem{qiao2023rotdiff}
H.~Qiao, S.~Feng, X.~Li, H.~Lin, H.~Hu, W.~Wei, and Y.~Ye, ``Rotdiff: A hyperbolic rotation representation model for information diffusion prediction,'' in \emph{CIKM}, 2023, pp. 2065--2074.

\bibitem{sankar2020inf}
A.~Sankar, X.~Zhang, A.~Krishnan, and J.~Han, ``Inf-vae: A variational autoencoder framework to integrate homophily and influence in diffusion prediction,'' in \emph{WSDM}, 2020, pp. 510--518.

\bibitem{hodas2014simple}
N.~O. Hodas and K.~Lerman, ``The simple rules of social contagion,'' \emph{Scientific reports}, vol.~4, no.~1, p. 4343, 2014.

\bibitem{zhong2012comsoc}
E.~Zhong, W.~Fan, J.~Wang, L.~Xiao, and Y.~Li, ``Comsoc: adaptive transfer of user behaviors over composite social network,'' in \emph{SIGKDD}, 2012, pp. 696--704.

\bibitem{li2017deepcas}
C.~Li, J.~Ma, X.~Guo, and Q.~Mei, ``Deepcas: An end-to-end predictor of information cascades,'' in \emph{WWW}, 2017, pp. 577--586.

\bibitem{cao2017deephawkes}
Q.~Cao, H.~Shen, K.~Cen, W.~Ouyang, and X.~Cheng, ``Deephawkes: Bridging the gap between prediction and understanding of information cascades,'' in \emph{CIKM}, 2017, pp. 1149--1158.

\bibitem{chen2019information}
X.~Chen, F.~Zhou, K.~Zhang, G.~Trajcevski, T.~Zhong, and F.~Zhang, ``Information diffusion prediction via recurrent cascades convolution,'' in \emph{ICDE}.\hskip 1em plus 0.5em minus 0.4em\relax IEEE, 2019, pp. 770--781.

\bibitem{wang2017topological}
J.~Wang, V.~W. Zheng, Z.~Liu, and K.~C.-C. Chang, ``Topological recurrent neural network for diffusion prediction,'' in \emph{ICDM}.\hskip 1em plus 0.5em minus 0.4em\relax IEEE, 2017, pp. 475--484.

\bibitem{yang2019neural}
C.~Yang, M.~Sun, H.~Liu, S.~Han, Z.~Liu, and H.~Luan, ``Neural diffusion model for microscopic cascade study,'' \emph{TKDE}, vol.~33, no.~3, pp. 1128--1139, 2019.

\bibitem{wang2022cascade}
D.~Wang, L.~Wei, C.~Yuan, Y.~Bao, W.~Zhou, X.~Zhu, and S.~Hu, ``Cascade-enhanced graph convolutional network for information diffusion prediction,'' in \emph{DASFAA}, 2022, pp. 615--631.

\bibitem{li2018learning}
D.~Li, Y.~Yang, Y.-Z. Song, and T.~Hospedales, ``Learning to generalize: Meta-learning for domain generalization,'' in \emph{AAAI}, vol.~32, no.~1, 2018.

\bibitem{li2021simple}
P.~Li, D.~Li, W.~Li, S.~Gong, Y.~Fu, and T.~M. Hospedales, ``A simple feature augmentation for domain generalization,'' in \emph{ICCV}, 2021, pp. 8886--8895.

\bibitem{ozyurt2023contrastive}
Y.~Ozyurt, S.~Feuerriegel, and C.~Zhang, ``Contrastive learning for unsupervised domain adaptation of time series,'' \emph{ICLR}, 2023.

\bibitem{li2023grass}
H.~Li, C.~Xia, T.~Wang, Z.~Wang, P.~Cui, and X.~Li, ``Grass: Learning spatial--temporal properties from chainlike cascade data for microscopic diffusion prediction,'' \emph{TNNLS}, 2023.

\bibitem{wang2023multiscale}
R.~Wang, X.~Xu, and Y.~Zhang, ``Multiscale information diffusion prediction with minimal substitution neural network,'' \emph{TNNLS}, 2023.

\bibitem{liu2021ttt++}
Y.~Liu, P.~Kothari, B.~Van~Delft, B.~Bellot-Gurlet, T.~Mordan, and A.~Alahi, ``Ttt++: When does self-supervised test-time training fail or thrive?'' \emph{NeurIPS}, vol.~34, pp. 21\,808--21\,820, 2021.

\bibitem{hatem2023point}
A.~Hatem, Y.~Qian, and Y.~Wang, ``Point-tta: Test-time adaptation for point cloud registration using multitask meta-auxiliary learning,'' in \emph{ICCV}, 2023, pp. 16\,494--16\,504.

\bibitem{wang2024heterogeneous}
Z.~Wang, H.~Huang, A.~Zheng, and R.~He, ``Heterogeneous test-time training for multi-modal person re-identification,'' in \emph{AAAI}, vol.~38, no.~6, 2024, pp. 5850--5858.

\bibitem{liu2023flood}
Y.~Liu, X.~Ao, F.~Feng, Y.~Ma, K.~Li, T.-S. Chua, and Q.~He, ``Flood: A flexible invariant learning framework for out-of-distribution generalization on graphs,'' in \emph{SIGKDD}, 2023, pp. 1548--1558.

\bibitem{zhang2024fully}
J.~Zhang, Y.~Wang, X.~Yang, and E.~Zhu, ``A fully test-time training framework for semi-supervised node classification on out-of-distribution graphs,'' \emph{TKDD}, 2024.

\bibitem{zhang2024t3rd}
H.~Zhang, X.~Liu, Q.~Yang, Y.~Yang, F.~Qi, S.~Qian, and C.~Xu, ``T3rd: Test-time training for rumor detection on social media,'' in \emph{WWW}, 2024, pp. 2407--2416.

\bibitem{sung2018learning}
F.~Sung, Y.~Yang, L.~Zhang, T.~Xiang, P.~H. Torr, and T.~M. Hospedales, ``Learning to compare: Relation network for few-shot learning,'' in \emph{CVPR}, 2018, pp. 1199--1208.

\bibitem{santoro2016meta}
A.~Santoro, S.~Bartunov, M.~Botvinick, D.~Wierstra, and T.~Lillicrap, ``Meta-learning with memory-augmented neural networks,'' in \emph{ICML}.\hskip 1em plus 0.5em minus 0.4em\relax PMLR, 2016, pp. 1842--1850.

\bibitem{finn2017model}
C.~Finn, P.~Abbeel, and S.~Levine, ``Model-agnostic meta-learning for fast adaptation of deep networks,'' in \emph{ICML}.\hskip 1em plus 0.5em minus 0.4em\relax PMLR, 2017, pp. 1126--1135.

\bibitem{ravi2017optimization}
S.~Ravi and H.~Larochelle, ``Optimization as a model for few-shot learning,'' in \emph{ICLR}, 2017.

\bibitem{sun2019meta}
Q.~Sun, Y.~Liu, T.-S. Chua, and B.~Schiele, ``Meta-transfer learning for few-shot learning,'' in \emph{CVPR}, 2019, pp. 403--412.

\bibitem{park2020fast}
S.~Park, J.~Yoo, D.~Cho, J.~Kim, and T.~H. Kim, ``Fast adaptation to super-resolution networks via meta-learning,'' in \emph{ECCV}.\hskip 1em plus 0.5em minus 0.4em\relax Springer, 2020, pp. 754--769.

\bibitem{soh2020meta}
J.~W. Soh, S.~Cho, and N.~I. Cho, ``Meta-transfer learning for zero-shot super-resolution,'' in \emph{CVPR}, 2020, pp. 3516--3525.

\bibitem{liu2019self}
S.~Liu, A.~Davison, and E.~Johns, ``Self-supervised generalisation with meta auxiliary learning,'' \emph{NeurIPS}, vol.~32, 2019.

\bibitem{gao2019popularity}
X.~Gao, Z.~Zheng, Q.~Chu, S.~Tang, G.~Chen, and Q.~Deng, ``Popularity prediction for single tweet based on heterogeneous bass model,'' \emph{TKDE}, vol.~33, no.~5, pp. 2165--2178, 2019.

\bibitem{cheng2014can}
J.~Cheng, L.~Adamic, P.~A. Dow, J.~M. Kleinberg, and J.~Leskovec, ``Can cascades be predicted?'' in \emph{WWW}, 2014, pp. 925--936.

\bibitem{shulman2016predictability}
B.~Shulman, A.~Sharma, and D.~Cosley, ``Predictability of popularity: Gaps between prediction and understanding,'' in \emph{ICWSM}, vol.~10, no.~1, 2016, pp. 348--357.

\bibitem{chen2020event}
X.~Chen, X.~Zhou, J.~Chan, L.~Chen, T.~Sellis, and Y.~Zhang, ``Event popularity prediction using influential hashtags from social media,'' \emph{TKDE}, vol.~34, no.~10, pp. 4797--4811, 2020.

\bibitem{mishra2016feature}
S.~Mishra, M.-A. Rizoiu, and L.~Xie, ``Feature driven and point process approaches for popularity prediction,'' in \emph{CIKM}, 2016, pp. 1069--1078.

\bibitem{tang2021fully}
X.~Tang, D.~Liao, W.~Huang, J.~Xu, L.~Zhu, and M.~Shen, ``Fully exploiting cascade graphs for real-time forwarding prediction,'' in \emph{AAAI}, vol.~35, no.~1, 2021, pp. 582--590.

\bibitem{cheng2024information}
Z.~Cheng, F.~Zhou, X.~Xu, K.~Zhang, G.~Trajcevski, T.~Zhong, and P.~S. Yu, ``Information cascade popularity prediction via probabilistic diffusion,'' \emph{TKDE}, 2024.

\bibitem{wang2019hierarchical}
Z.~Wang and W.~Li, ``Hierarchical diffusion attention network,'' in \emph{IJCAI}, 2019, pp. 3828--3834.

\bibitem{koch2015siamese}
G.~Koch, R.~Zemel, R.~Salakhutdinov \emph{et~al.}, ``Siamese neural networks for one-shot image recognition,'' in \emph{ICML deep learning workshop}, vol.~2, no.~1.\hskip 1em plus 0.5em minus 0.4em\relax Lille, 2015, pp. 1--30.

\end{thebibliography}

\section{Biography Section}
 
\vspace{-0.4em} 

\begin{IEEEbiography}[{\includegraphics[width=1in,height=1.25in,clip,keepaspectratio]{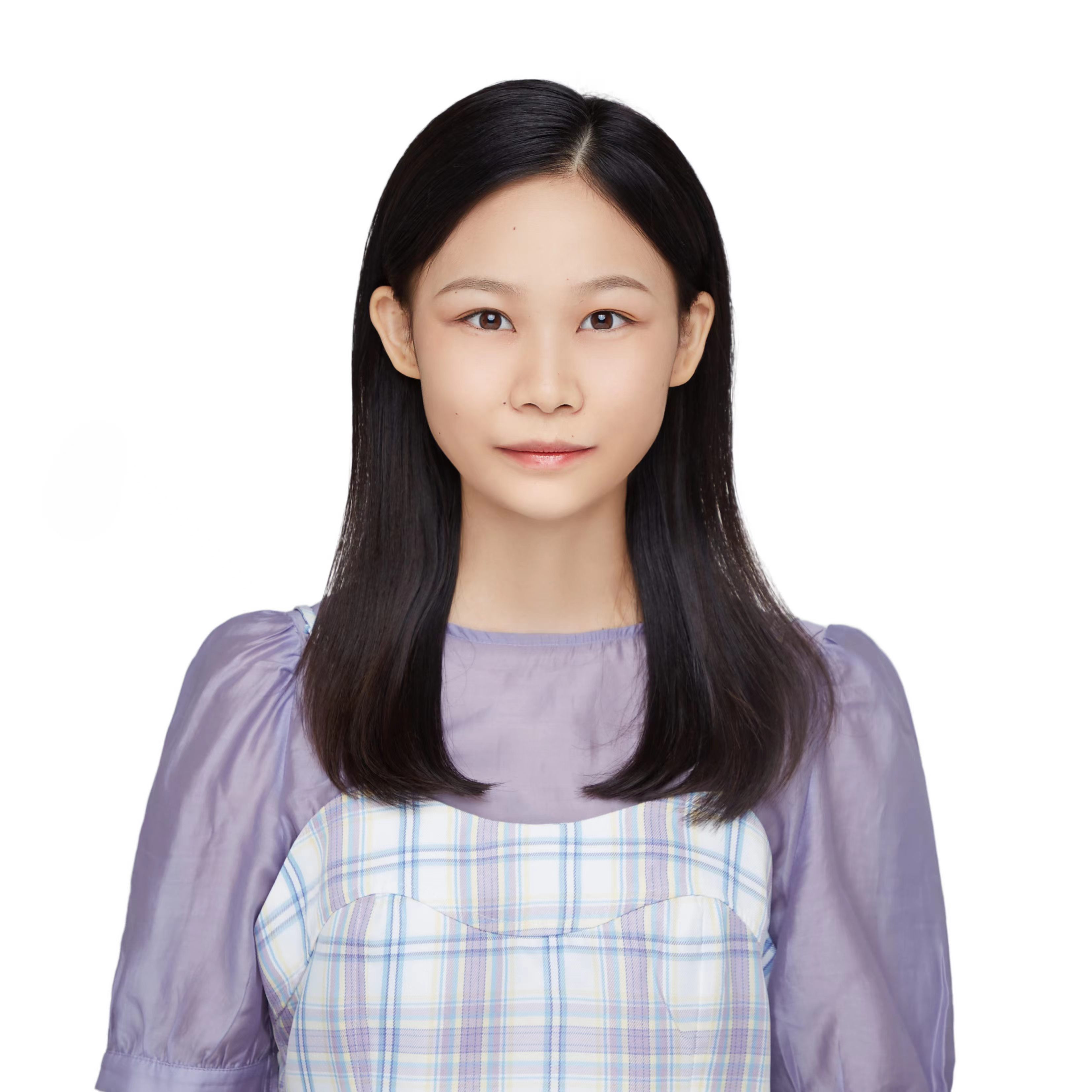}}]{Wenting Zhu} received the B.S. degree in Computer Science and Technology from Shandong University, China, in 2021. She is currently pursuing the Ph.D. degree with the Key Laboratory of Trustworthy Distributed Computing and Services, Ministry of Education, Beijing University of Posts and Telecommunications, China. Her research interests include information diffusion prediction and social network analysis.
\end{IEEEbiography}
\vspace{-0.4em} 

\begin{IEEEbiography}[{\includegraphics[width=1in,height=1.25in,clip,keepaspectratio]{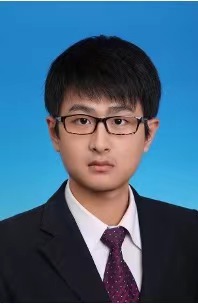}}]{ChaoZhuo Li} received the Ph.D. degree in computer software and theory from Beihang University, China, in 2020. From 2017 to 2019, he was a joint Ph.D. student at the University of Illinois at Urbana-Champaign, USA. From 2020 to 2024, he was a Senior Researcher with Microsoft Research Asia. He is currently an Associate Professor at the Beijing University of Posts and Telecommunications, China. He has published more than 100 research papers in top-tier venues such as NeurIPS, AAAI, SIGIR, ICDM, and CIKM. His research interests include large language model security, natural language processing, and social network analysis.
\end{IEEEbiography}

\vspace{-0.4em} 

\begin{IEEEbiography}[{\includegraphics[width=1in,height=1.25in,clip,keepaspectratio]{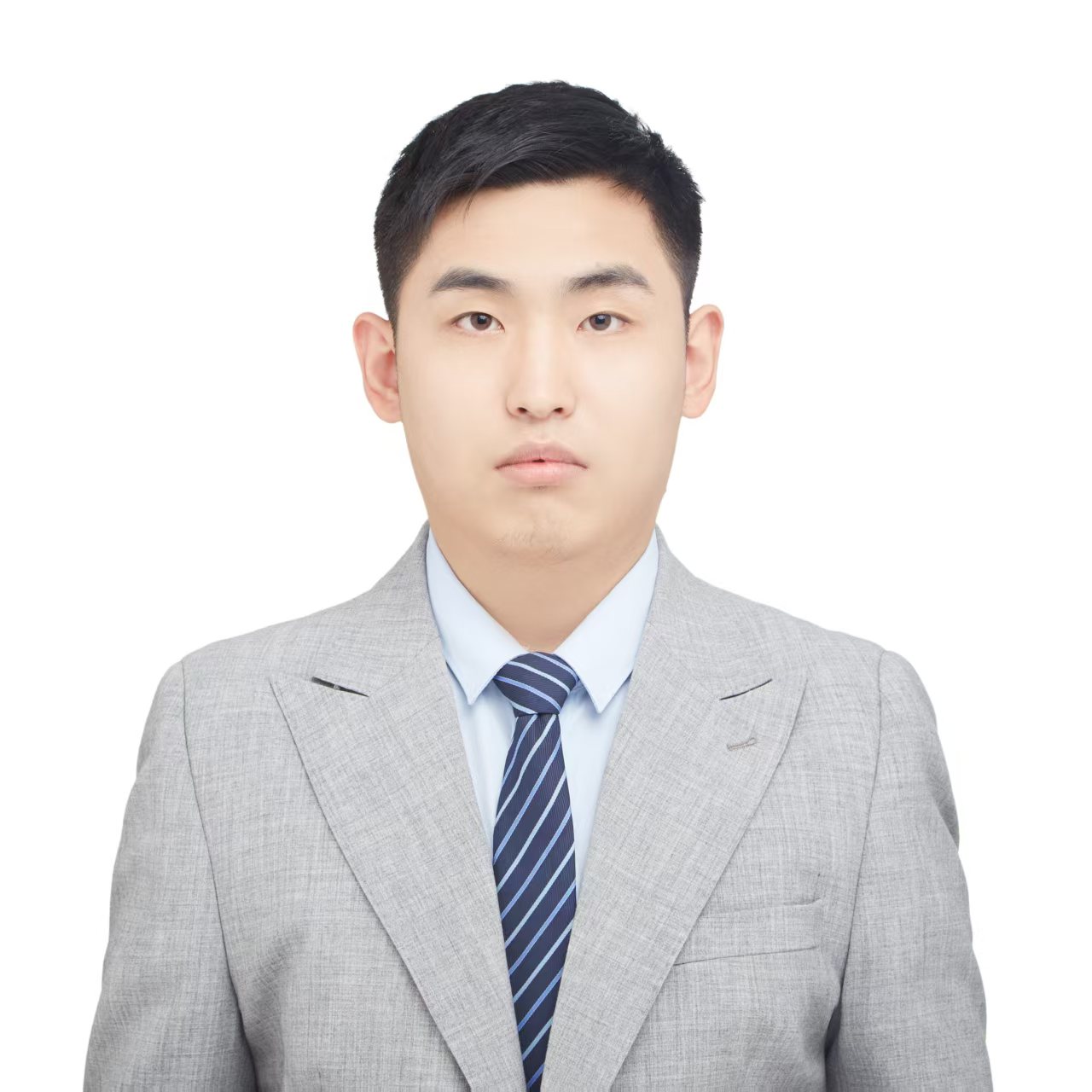}}]{Qingpo Yang} received the B.S. degree in Computer Science and Technology from North China Electric Power University, China, in 2022. He is currently pursuing the M.S. degree at the Key Laboratory of Trustworthy Distributed Computing and Services, Ministry of Education, Beijing University of Posts and Telecommunications, China. His research interests include cyberbullying detection.
\end{IEEEbiography}

\vspace{-0.4em} 

\begin{IEEEbiography}[{\includegraphics[width=1in,height=1.25in,clip,keepaspectratio]{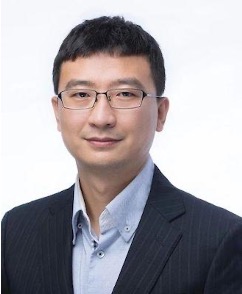}}]{Xi Zhang}(Member, IEEE) received the Ph.D. degree in computer science from Tsinghua University, China. He is a Professor at the Beijing University of Posts and Telecommunications, where he serves as the Vice Dean of the School of Cyberspace Security and the Vice Director of the Key Laboratory of Trustworthy Distributed Computing and Services, Ministry of Education, China. He is also a recipient of a national-level youth talent program in China. He was a visiting scholar at the University of Illinois at Chicago from 2015 to 2016. His research interests include information diffusion games, misinformation and harmful content detection, large language model security, and interpretable machine learning.
\end{IEEEbiography}

\vspace{-0.4em} 

\begin{IEEEbiography}[{\includegraphics[width=1in,height=1.25in,clip,keepaspectratio]{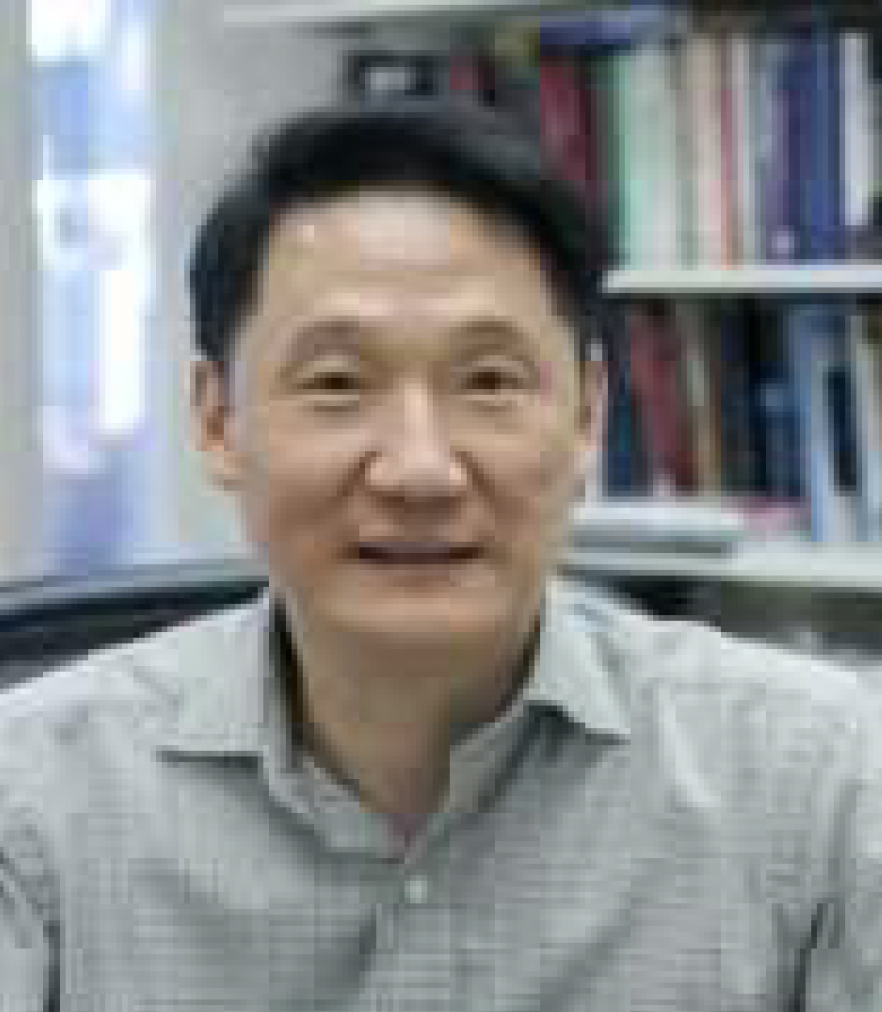}}]{Philip S. Yu}(Fellow, IEEE) received the B.S. Degree in E.E.from National Taiwan University, the M.S. and Ph.D. degrees in E.E. from Stanford University, and the M.B.A. degree from New York University. He is a Distinguished Professor in Computer Science at the University of Illinois at Chicago and also holds the Wexler Chair in Information Technology. Before joining UlC, Dr. Yu was with IBM, where he was manager of the Software Tools and Techniques department at the Watson Research Center. His research interest is on big data, including data mining, data stream, database, and privacy. He has published more than 1.500 papers in refereed journals and conferences. He holds or has applied for more than 300 US patents. Dr. Yu is a Fellow of the ACM and the IEEE. Dr. Yu is the recipient of the ACM SIGKDD 2016 Innovation Award for his influential research and scientific contributions on mining, fusion, and anonymization of big data. He also received the VLDB2022 Test of Time Award, ACM SIGSPATIAL 2021 10-year impact Award, and the EDBT 2014 Test of Time Award. He was the Editor-in-Chiefs of ACM Transactions on Knowledge Discovery from Data (2011-2017) and IEEE Transactions on Knowledge and Data Engineering (2001-2004).
\end{IEEEbiography}

\bibliographystyle{IEEEtran}

\vfill

\end{document}